\documentclass[twocolumn,twocolappendix]{aastex631}
\pdfoutput=1
\usepackage{booktabs}
\usepackage{amsmath}
\usepackage{csquotes}

\shorttitle{COSMOS2020: LSS and Galaxy Environments}
\shortauthors{Taamoli et al.}

\begin{document}

\title{Large Scale Structures in COSMOS2020: Evolution of Star Formation Activity in Different Environments at $0.4 < z < 4$}

\author[0000-0003-0749-4667]{Sina Taamoli}
\affiliation{Department of Physics and Astronomy, University of California, Riverside, 900 University Ave, Riverside, CA 92521, USA}

\author[0000-0001-5846-4404]{Bahram Mobasher}
\affiliation{Department of Physics and Astronomy, University of California, Riverside, 900 University Ave, Riverside, CA 92521, USA}

\author[0000-0003-3691-937X]{Nima Chartab}
\affiliation{Observatories of the Carnegie Institution of Washington: Pasadena, CA, USA}

\author[0000-0003-4919-9017]{Behnam Darvish}
\affiliation{Department of Physics and Astronomy, University of California, Riverside, 900 University Ave, Riverside, CA 92521, USA}

\author[0000-0003-1614-196X]{John R. Weaver}
\affiliation{Department of Astronomy, University of Massachusetts, Amherst, MA 01003, USA}

\author[0000-0003-2226-5395]{Shoubaneh Hemmati}
\affiliation{Infrared Processing and Analysis Center, California Institute of Technology, Pasadena, CA 91125, USA}

\author[0000-0002-0930-6466]{Caitlin M. Casey}
\affiliation{Department of Astronomy, The University of Texas at Austin, 2515 Speedway Boulevard Stop C1400, Austin, TX 78712, USA}

\author[0000-0002-0364-1159]{Zahra Sattari}
\affiliation{Department of Physics and Astronomy, University of California, Riverside, 900 University Ave, Riverside, CA 92521, USA} 
\affiliation{Observatories of the Carnegie Institution of Washington: Pasadena, CA, USA}

\author[0000-0003-2680-005X]{Gabriel Brammer}
\affiliation{Cosmic Dawn Center (DAWN), Denmark}
\affiliation{Niels Bohr Institute, University of Copenhagen, Jagtvej 128, DK-2200 Copenhagen, Denmark}

\author[0000-0003-3578-6843]{Peter L. Capak}
\affiliation{Infrared Processing and Analysis Center, California Institute of Technology, Pasadena, CA 91125, USA}

\author[0000-0002-7303-4397]{Olivier Ilbert}
\affiliation{Aix Marseille Univ, CNRS, LAM, Laboratoire d’Astrophysique de Marseille, Marseille, France}

\author[0000-0001-9187-3605]{Jeyhan S. Kartaltepe}
\affiliation{Laboratory for Multiwavelength Astrophysics, School of Physics and Astronomy, Rochester Institute of Technology, 84 Lomb Memorial Drive, Rochester, NY
14623, USA}

\author[0000-0002-9489-7765]{Henry J. McCracken}
\affiliation{Institut d’Astrophysique de Paris, UMR 7095, CNRS, and Sorbonne Université, 98 bis boulevard Arago, F-75014 Paris, France}

\author{Andrea Moneti}
\affiliation{Institut d’Astrophysique de Paris, UMR 7095, CNRS, and Sorbonne Université, 98 bis boulevard Arago, F-75014 Paris, France}

\author[0000-0002-1233-9998]{David B. Sanders}
\affiliation{Institute for Astronomy (IfA), University of Hawaii, 2680 Woodlawn Drive, Honolulu, HI 96822, USA}

\author[0000-0002-0438-3323]{Nicholas Scoville}
\affiliation{California Institute of Technology, 1200 E. California Boulevard, Pasadena, CA 91125, USA}

\author[0000-0003-3780-6801]{Charles L. Steinhardt}
\affiliation{Cosmic Dawn Center (DAWN), Denmark}
\affiliation{Niels Bohr Institute, University of Copenhagen, Lyngbyvej 2, DK-2100 Copenhagen Ø, Denmark}

\author[0000-0003-3631-7176]{Sune Toft}
\affiliation{Cosmic Dawn Center (DAWN), Denmark}
\affiliation{Niels Bohr Institute, University of Copenhagen, Jagtvej 128, DK-2200 Copenhagen, Denmark}

\begin{abstract}
To study the role of environment in galaxy evolution, we reconstruct the underlying density field of galaxies based on COSMOS2020 (The Farmer catalog) and provide the density catalog for a magnitude limited ($K_{s}<24.5$) sample of $\sim 210 \, k$ galaxies at $0.4<z<5$ within the COSMOS field. The environmental densities are calculated using weighted Kernel Density Estimation (wKDE) approach with the choice of von Mises-Fisher kernel, an analog of the Gaussian kernel for periodic data. Additionally, we make corrections for the edge effect and masked regions in the field. We utilize physical properties extracted by LePhare to investigate the connection between star formation activity and the environmental density of galaxies in six mass-complete sub-samples at different cosmic epochs within $0.4<z<4$. Our findings confirm a strong anti-correlation between star formation rate (SFR)/specific SFR (sSFR) and environmental density out to $z \sim 1.1$. At $1.1<z<2$, there is no significant correlation between SFR/sSFR and density. At $2<z<4$ we observe a reversal of the SFR/sSFR-density relation such that both SFR and sSFR increase by a factor of $\sim 10$ with increasing density contrast, $\delta$, from -0.4 to 5. This observed reversal at higher redshifts supports the scenario where an increased availability of gas supply, along with tidal interactions and a generally higher star formation efficiency in dense environments, could potentially enhance star formation activity in galaxies located in rich environments at $z>2$.
\end{abstract}


\section{Introduction}\label{sec:intro}
Galaxies in the universe are distributed in a web-like structure known as \enquote{Cosmic Web} \citep{bond1996}. The study of these large-scale structures (hereafter LSS), which comprise galaxy clusters, sparsely populated voids, filamentary threads, and planar walls, is a cornerstone in our understanding of the evolution of galaxies and dark matter, which are strongly connected.

The identification of LSS and the study of matter distribution within the cosmic web is still challenging due to the diverse shape and size of LSS components, which often confines such studies to local-universe spectroscopic surveys \citep{york2000,colless2001}, simulations \citep{cautun2014,Vogelsberger2014,libeskind2018}, and analytical methods \citep{Bardeen1986,bond1996,Sousbie2011,ansari22}. However, with the advent of wide and deep photometric surveys using ground and space telescopes, such as The Cosmic Evolution Survey (COSMOS), we are now able to identify and study these structures and their impact on the galaxy evolution in the high-redshift universe. Several studies have confirmed and investigated LSS in the COSMOS field, including the following examples: identification of 247 X-ray groups at $0.08<z<1.53$ \citep{Gozaliasl2019}, study of a large filamentary structure, known as COSMOS wall, at $z\sim0.73$ \citep{iovino2016}, spectroscopic confirmation/investigation of a large scale structure at $z\sim2.1$ \citep{hung_large-scale_2016}, a protocluster at $z\sim2.23$ \citep{darvish_spectroscopic_2020}, study of complex-shaped overdensities at $z\sim2.45$ using photometric and spectroscopic observations \citep{cucciati_progeny_2018}, an asymmetric filamentary structure at $z\sim2.47$ \citep{casey_massive_2015}, a concentrated group of massive galaxies with extended X-ray emission at $z\sim 2.506$ \citep{wang_discovery_2016}, a protocluster of massive quiescent galaxies at $z\sim2.77$ \citep{ito2023}, massive protoclusters at $z\sim3.3$ \citep{forrest_elentarimassive_2023}, $z\sim 3.366$ \citep{mcconachie_spectroscopic_2022}, $z\sim4.57$ \citep{lemaux_vimos_2018}, a dense group with a spectroscopically confirmed quiescent galaxy at its center at $z\sim 4.53$ \citep{Kakimoto2023}, and $z\sim 5.3$ \citep{capak_massive_2011}, and discovery of a massive, dusty starburst galaxy in a protocluster at $z\sim5.7$ \citep{pavesi_hidden_2018}. Additionally, overdensities in 3D Ly$\alpha$ forest tomography are studied as alternative tracers of LSS (e.g., CLAMATO, \citealt{lee_shadow_2016} and LATIS, \citealt{newman_latis_2020} surveys) and the evolution of previously identified protoclusters over $\sim 11 \, \text{Gyr}$ is studied through constrained simulations in \citep{Ata2022}.

Alongside the identification of LSS in spectroscopic and photometric samples, there has been a notable increase in the study of galaxy properties across different environments in recent decades. This includes several studies focusing on the evolution of morphology \citep{Mandelbaum2006,capak2007,Bamford2009}, gas content \citep{Catinella2013}, star formation activity \citep{Scoville2013,darvish2016,chartab2020}, and quenching mechanisms \citep{peng2010,Poggianti2017,zheng2024} in different environments.

Studies show that in the local universe, early-type passive galaxies are typically found in denser environments, such as galaxy clusters, while late-type star-forming galaxies are mainly located in less-dense regions, known as field \citep{dressler1980,balogh2004,kauffmann2004,peng2010,woo2013,baldry2006}. This is partly because in addition to internal processes, such as gas outflows due to supernova explosions, stellar and active galactic nuclei (AGN) feedback \citep{dekel1986,dallavecchia2008,fabian2012,Bremer2018}, galaxies in denser environments have experienced an enhanced level of \enquote{environmental quenching} mechanisms such as ram pressure stripping \citep{gunn_infall_1972,Moore1999,Brown2017,Barsanti2018}, strangulation or
starvation \citep{Moore1999,peng2015}, galaxy harassment \citep{moore_galaxy_1996,farouki_computer_1981}.

At higher redshifts (out to $z \sim 1.4$), \citep{capak2007} investigated the density-morphology relations, finding that galaxies are transformed from late (spiral and irregular) to early-type galaxies more rapidly in dense regions compared to sparse regions.

While these trends are well-established in the lower redshifts, they remain a matter of ongoing debate at intermediate and higher redshifts ($z\gtrsim1$). \citep{patel2009} reports a negative correlation between star formation activity and environmental density at $z\sim0.83$, the same as the local universe. Using a large sample of galaxies in 5 CANDELS fields (GOODSN, GOODS-S, EGS, UDS, and COSMOS), \citep{chartab2020} further extended this observation to as high as $z\sim 3.5$. There are also studies that find no significant correlation between SFR and environment beyond redshift $z\sim 1$ \citep{Grutzbauch2011, Scoville2013, darvish2016}. Conversely, there are indications that the low redshift trends between SFR and environmental density begin to diminish, or even reverse, around $z\sim1-2$. This potential reversal has been observed in several studies that examine galaxies within individual structures and compare their star formation activity with counterparts in field environments. For example, the reversal of this trend at $z\sim 0.8-1$ was reported using a sample of galaxies in GOODS field by \citep{elbaz2007} and a sample drawn from SDSS and DEEP2 redshift survey by \citep{cooper2008}. Utilizing a large sample of spectroscopically confirmed galaxies, \citep{lemaux_vimos_2022} reported a monotonic increase in SFR with increasing galaxy overdensity in the early universe ($2<z<5$). However, some studies attribute the observed reversal to cosmic variance, AGN contamination, and various dynamical ranges of environments used in different studies \citep{sobral2011,Scoville2013,darvish2016}. In any case, the discrepancy between high redshift results is due to the limited availability of observational data at higher redshifts, a lack of complete samples of statistically significant size, and uncertainties in the extracted physical parameters for fainter objects. This weakens the statistical reliability of the observed trends at higher redshifts and makes the interpretation of these observations more challenging. Thus, more studies are needed to reliably identify the LSS at higher redshifts and investigate the role of the environment in the star formation activity of galaxies in the early universe.

COSMOS data spans a large area of ($\sim2 \, \text{deg}^{2}$) which enables us to investigate these correlations in a variety of environments with potentially lower impacts of cosmic variance on the results. It is important to note that COSMOS does not appear to have many massive structures at ($z\lesssim 2$), and the dynamic range of overdensities at low redshifts is fairly small when compared to the other local-universe surveys (SDSS, \citealt{york2000}) or those that specifically target fields that contain massive LSS at $z\sim1$ (e.g., EdisCS, \citealt{White2005}; GOGREEN, \citealt{Balogh2017}; ORELSE, \citealt{lubin2009}).  Nonetheless, the latest release of the COSMOS catalog, COSMOS2020, with its deeper optical, infrared, and near-infrared data compared to previous releases, offers an opportunity to extend studies of LSS and its impact on galaxy evolution to higher redshifts. Since its publication, COSMOS2020 catalog has been used in several extragalactic studies (e.g., \citealt{ito2022}, \citealt{shuntov2022}, \citealt{davidzon2022}, \citealt{kauffmann2022}, \citealt{Gould2023}, \citealt{Toni2023}).
 
To start our analysis, we first need to clarify what we mean by \enquote{environmental density}. Numerous methods have been used in the literature to estimate the density field associated with a given distribution of galaxies. A comprehensive review and comparison between these methods including \enquote{weighted Kernel Density Estimation (wKDE)}, \enquote{weighted K-Nearest Neighbor}, \enquote{weighted Voronoi Tesselation}, and \enquote{weighted Delaunay Triangulation} is provided by \citep{darvish2015}. Examining the performance of all these methods on simulated data, \citep{darvish2015} conclude that the weighted Kernel Density (wKDE) and Voronoi tesselation best reproduce the underlying density field in simulated data and result in a lower mean squared error (MSE) when applied on simulated data. According to \citep{darvish2015}, wKDE is suitable for weighted data and less affected by the shot noise, which becomes important in sparse distributions, and possible random clustering of foreground and background sources.

In this study, we adopt wKDE method with the choice of von Mises-Fisher kernel function to estimate the underlying density field at different redshifts using COSMOS2020 catalog. We produce density maps within the $0.4<z<6$ range which can be used to identify LSS and release a publicly available catalog of measured densities for 210621 galaxies brighter than $K_{s} < 24.5$ at $0.4 < z < 5$ in the COSMOS field. We implement corrections to mitigate edge effects and masked sources in the vicinity of bright stars to improve the quality of density estimation. Eventually, we investigate the relation between estimated environmental density and the star formation activity (SFR/sSFR) at different redshift intervals to study the evolution of this relation with cosmic time. 

The paper is organized as follows: In Section \ref{sec:data}, we introduce the properties of data and selection criteria we used in this study. In Section \ref{sec:density}, we describe the method used to construct density maps and the environmental density catalog, and we present and discuss our results in Section \ref{sec:results}. We summarize our findings in Section \ref{sec:Summary}.

Throughout this work, we assume a flat $\Lambda\text{CDM}$ cosmology with $H_{0} = 70 \, \text{km} \, s^{-1} \text{Mpc}^{-1}$ , $\Omega_{m0} = 0.3$, and $\Omega_{\Lambda0}= 0.7 $. All magnitudes are expressed in the AB system and the physical parameters are measured assuming a Chabrier initial mass function.

\section{Data} \label{sec:data}
COSMOS2020 consists of $\sim1.7$ million sources for which source detection and multiwavelength photometry (X-ray to radio-imaging) are performed across $\sim 2 \, \text{deg}^{2}$ of the equatorial COSMOS field \citep{weaver2022}.  In addition to the previous release of this catalog (COSMOS2015, \citealt{Laigle2016}), COSMOS2020 consists of new ultra-deep optical data from the Hyper Supreme-Cam (HSC), Subaru Strategic Program (SSP), and new Visible Infrared Survey Telescope for Astronomy (VISTA) data from DR4 reaching more than one magnitude deeper in the $K_{s}$ band over the full area. Deep $u^{\star}$- and new $u$-band imaging from the Canada-France-Hawaii Telescope program CLAUDS \citep{Sawicki2019} provides us with a deep coverage over a greater area than COSMOS2015.

In COSMOS2020, around $966000$ sources are measured with all available broadband data utilizing two photometry tools 1) traditional aperture photometry \citep{Laigle2016}, the \enquote{\textit{CLASSIC}} catalog, and 2) a new profile-fitting photometric tool, \enquote{\texttt{The Farmer}} (\citealt{weaver_farmer_2023}a, \citealt{farmer2023}b), each of which includes photometric redshifts (hereafter photo-$z$) and other physical parameters computed by \texttt{EASY} \citep{brammer2008easy} and \texttt{LePhare}\footnote{\href{https://www.cfht.hawaii.edu/~arnouts/LEPHARE/lephare.html}{www.cfht.hawaii.edu/\(\sim \)arnouts/LEPHARE/lephare.html}} \citep{ilbert2006, Arnouts2002lephare}. Both \texttt{The Farmer} and \textit{CLASSIC} photometries are corrected for dust extinction using the \citep{Schlafly2011} dust map. In this study, we use photo-$z$s and other physical parameters derived by the combination of \texttt{The Farmer} photometric measurements and \texttt{LePhare} SED fitting code.

\texttt{The Farmer} is a profile-fitting photometry package that combines a library of smooth parametric models from The Tractor \citep{Lang2016Tractor} with a decision tree that aims to determine the best-fit model in harmony with neighboring (blended) sources. The resulting photometric measurements are naturally total, without the need for aperture corrections, and more reliable in deep extragalactic fields with more crowded regions \citep{weaver2022}. According to \citep{weaver_farmer_2023}, \texttt{The Farmer} is particularly effective at de-blending sources in low-resolution images like IRAC. In contrast, aperture photometry (the \textit{CLASSIC} catalog) may underestimate the total flux of the sources and does not simultaneously model blended objects. Overall, the photo-$z$ quality is similar between the two catalogs but, as noted by \citep{weaver2022}, the Farmer demonstrates superior performance at fainter magnitudes ($i\gtrsim 24$) which predominantly correspond to high-redshift sources. With our choice of magnitude cut $24.5$ (AB) on $K_{s}$ band, our magnitude limited sample remains complete down to $\sim 24.9$ (AB) in $i$-band and $\sim 32\%$ of sources fall within the range of $24\lesssim i \lesssim 24.9$, where \texttt{Farmer} demonstrates superior performance compared to the \textit{CLASSIC} catalog. Better performance at fainter magnitudes and its more effective modeling of blended sources in densely populated regions led us to choose \texttt{The Farmer} catalog for this study. However, the drawback of this choice is that Farmer fails to model sources in the vicinity of bright sources, excluding $\sim 18\%$ of sources in masked regions. Two catalogs are compared in full detail in \citep{weaver2022}.

\begin{figure}
    \centering
    \includegraphics[width = 0.95\linewidth]{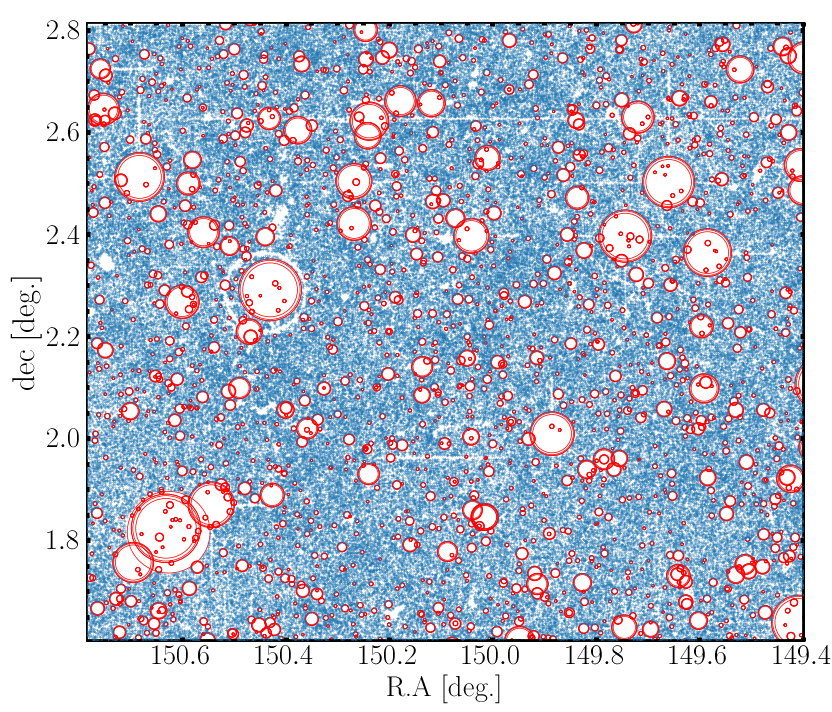}
    \caption{Spatial distribution of sources selected based on criteria outlined in Section \ref{sec:data}. Red circles show bright star masks from the HSC-SSP PDR2 ($\sim 7600$  regions) provided by \citep{coupon2018} which are used to mask objects in the vicinity of bright stars.}
    \label{cosmos}
\end{figure}

For the physical parameters, we use the results extracted by \texttt{LePhare} \citep{Arnouts2002lephare,ilbert2006} which uses the same configuration outlined in \citep{ilbert2013} to fit both galaxy and stellar templates to the observed photometry. As the first step, photo-$z$s are estimated following the method outlines in \citep{Laigle2016}, then the physical properties such as absolute magnitudes, star formation rates (SFR/sSFR), and stellar mass are computed with the same configuration as COSMOS2015: \texttt{LePhare} fits a template library generated by \citep{Bruzual2003} models to the observed photometry after fixing the redshift of each target to the estimated photo-$z$ in the first step. Further details are discussed in \citep{Laigle2016,weaver2022}.

The catalog contains photometric measurements in 44 bands (including U-band, Optical, Near-infrared, Mid-infrared, X-ray, UV, and HST data), area flags, object type, and physical parameters such as SFR, sSFR, and stellar mass. Throughout this work, we use the median values of the photo-$z$ probability distribution function (zPDF) and the marginalized likelihood of other physical parameters such as stellar mass, SFR, and sSFR, as reported in COSMOS2020. We limit our study to a sub-sample of 211431 galaxies with the following selection criteria:

\begin{itemize}
    \item Photo-$z$ range of $0.4<z<6$: although we report environmental densities for sources out to $z\sim5$ and the primary focus of our star formation activity-environment analysis is on sources within $0.4 < z < 4$, we considered a buffer range ($4<z<6$) at the high redshift end of our primary redshift range to capture the full (zPDF) of sources whose zPDFs extend tails beyond $z=4$. Since zPDFs are narrower for low-redshift sources, there is no need for such a buffer range at the lower end of our redshift range. Because of the sparsity of sources and a bias toward brighter sources, the reliability of the reconstructed density fields significantly diminishes beyond $z \sim 4.5-5$.
    \item We filter out sources with large uncertainties in their photo-$z$ measurements ($\Delta z > 2$). $\Delta z$ is the 68\% confidence interval on estimated photo-$z$. These sources do not effectively contribute to the density field. The relative median redshift uncertainty of the filtered sample is $\Delta z / (1+z)\lesssim0.02$ throughout the entire redshift range.
    \item An area of $\sim 1.7 \, deg^{2}$ enclosed within $1.604<\delta<2.817$ and $149.398<\alpha<150.787$. This is the largest region containing robust NIR imaging equating to a spatially homogeneous selection function and NIR coverage.
    \item \texttt{LePhare} separates galaxies from stars and AGNs by combining morphological and SED criteria. We use a pure sample of \enquote{galaxies} (as identified by \texttt{LePhare}) that are not in the \enquote{bright star} masks.
    \item Magnitude cut of value $24.5$ (AB) mag on Ultra-Vista $K_{s}$, as the photo-$z$ uncertainties increase significantly for fainter sources.
\end{itemize}

\begin{figure}
    \centering
    \includegraphics[width = 1\linewidth]{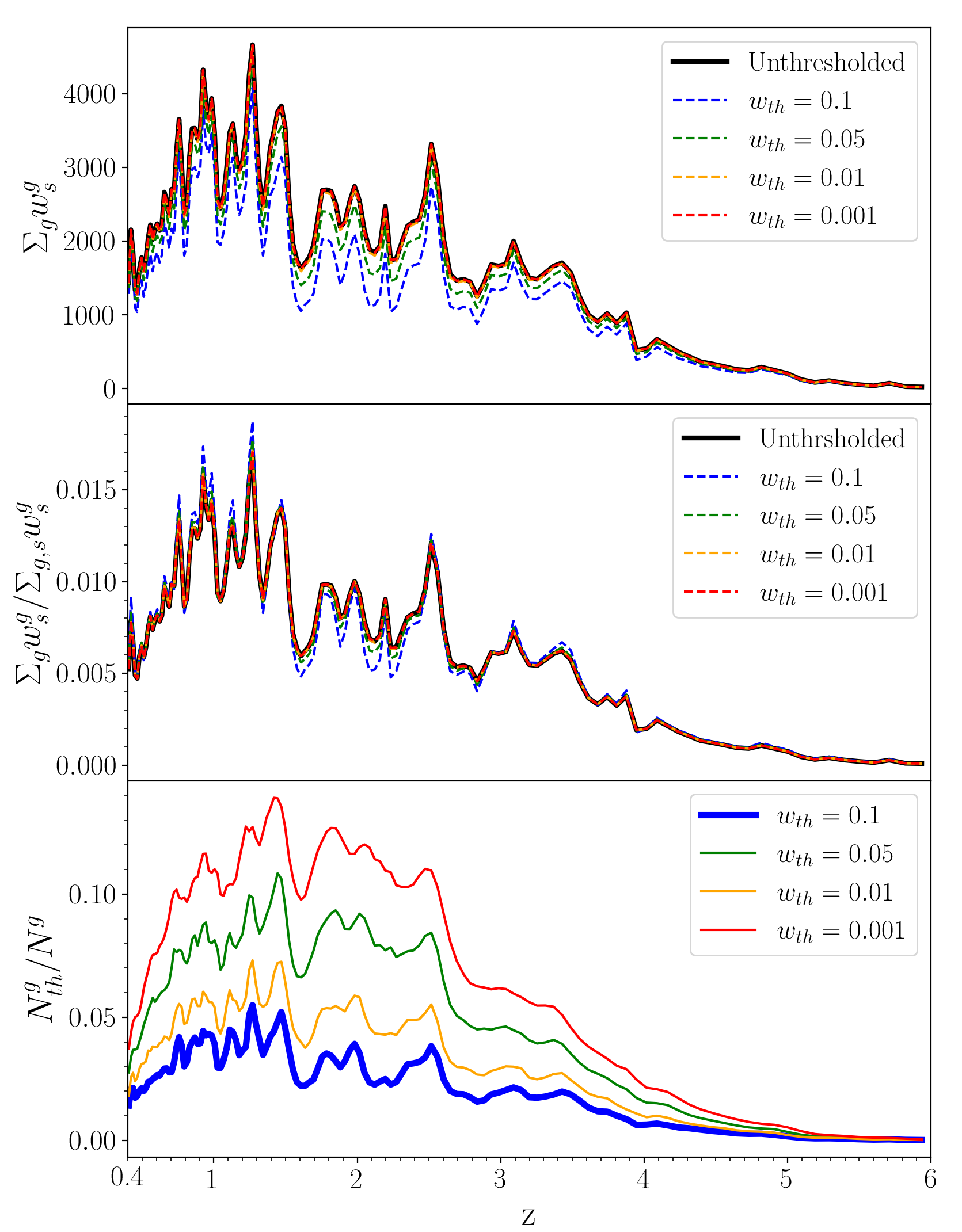}
    \caption{Top: the effective number of galaxies ($\sum_{g} w_{s}^{g}$) in each slice, for different thresholds on weight: 0.1, 0.05, 0.01, 0.001. The black curve shows the full sample without a threshold on weights. Middle: the effective number of galaxies in each redshift slice divided by the effective number of galaxies in all slices. Bottom: the size of the thresholded sample divided by the size of the unthresholded sample in each redshift slice. This fraction shows the degree to which thresholding lessens the computational time in the density estimate stage.}
    \label{weights}
\end{figure}

Figure \ref{cosmos} shows the spatial distribution of the resulting sample across the field. Red circles show $\sim 7600$ masked regions that contain sources in the vicinity of bright stars \citep{coupon2018}.

\section{Density Field Estimation} \label{sec:density}
We calculate the environmental densities adopting the same approach introduced in \cite{chartab2020} with minor modifications detailed in the following sections. Although we refer the reader to \cite{chartab2020} for more comprehensive details, we summarize the key steps of the method in the following section for clarity, and to contextualize the modifications we implemented. 

\subsection{Weighted Kernel Density Estimation}
wKDE is a non-parametric method used for density reconstruction based on the spatial distribution of data points \citep{parzen1962,Guillamón1998,Gisbert2003,darvish2015}, and is especially effective in handling weighted data. To reconstruct the density field utilizing wKDE, we implement the following procedure: 1) dividing the sample into redshift slices (Section \ref{sec: zslice}), 2) calculating weights for all galaxies in each redshift slice (Section \ref{sec: weightcalculation}), 3) applying corrections to improve the density estimation around the masked regions (Section \ref{sec:maskedregions}), 4) find the optimum bandwidth in each redshift slice and an adaptive bandwidth for each source (Section \ref{sec:bandwidthselection}), 5) applying corrections for the \enquote{edge effect} that impacts the estimated density map near the edges of the field (Section \ref{sec:boundaryeffect}).

In this scheme, the surface density for the $i$th galaxy in the $s$th redshift slice and at position ($X_{i}$) is defined as:
\begin{equation}\label{eq:densitySlice}
    \sigma^{s}(X_{i}) = \sum_{g}w_{g}^{s}K(X_{i};X_{g})
\end{equation}
where, $K(X_{i}; X_{g})$, is the Kernel function of our choice calculated between two sources at $X_{i}$ and $X_{g}$ and $w_{g}^{s}$ is the probability of galaxy $g$ being at $s$th redshift slice normalized to the sum of its weights in all redshift slices. In this work, we adopt the \enquote{von Mises-Fisher} kernel function, which is the analog of the Gaussian kernel function for circular/periodic data (here R.A and Dec. coordinates) \citep{bai_kernel_1988,GARCIAPORTUGUES2013152,taylor_2008,chartab2020}. The \enquote{von Mises-Fisher} kernel is a simplified, isotropic form of a more general 5-parameter kernel function known as \enquote{Kent} distribution \citep{kent1982}. While the \enquote{von Mises-Fisher} kernel effectively reduces to the Gaussian kernel in small fields, its use is particularly advantageous for providing greater accuracy in future wide-field surveys: 
\begin{equation}
    K(X_{i};X_{g}) = \frac{1}{4\pi b^{2} \sinh (1/ b^{2})}\exp \left( \frac{\cos \psi}{b^{2}}\right)
\end{equation}
Where $\psi$ is the angular distance between $X_{i}$ and $X_{g}$ which can be calculated using their coordinates and $b$ is the bandwidth of this kernel which determines the extent to which the source at position $X_{g}$ contributes to the environmental density of the source at position $X_{i}$. Details of finding the optimum value of $b$ are explained in Section \ref{sec:bandwidthselection}. 

Eventually, We calculate the environmental density for the $i$th galaxy in our sample, $\sigma(G_{i})$, as a weighted sum of the surface density across all redshift slices \citep{chartab2020}:
\begin{equation}
    \sigma (G_{i}) = \sum_{s}w_{i}^{s} \sigma_{s}(X_{i})
\end{equation}
Where $w_{i}^{s}$ is the weight associated with the galaxy $i$ at $s$th slice.

\begin{figure}
    \centering
    \includegraphics[width = 1\linewidth]{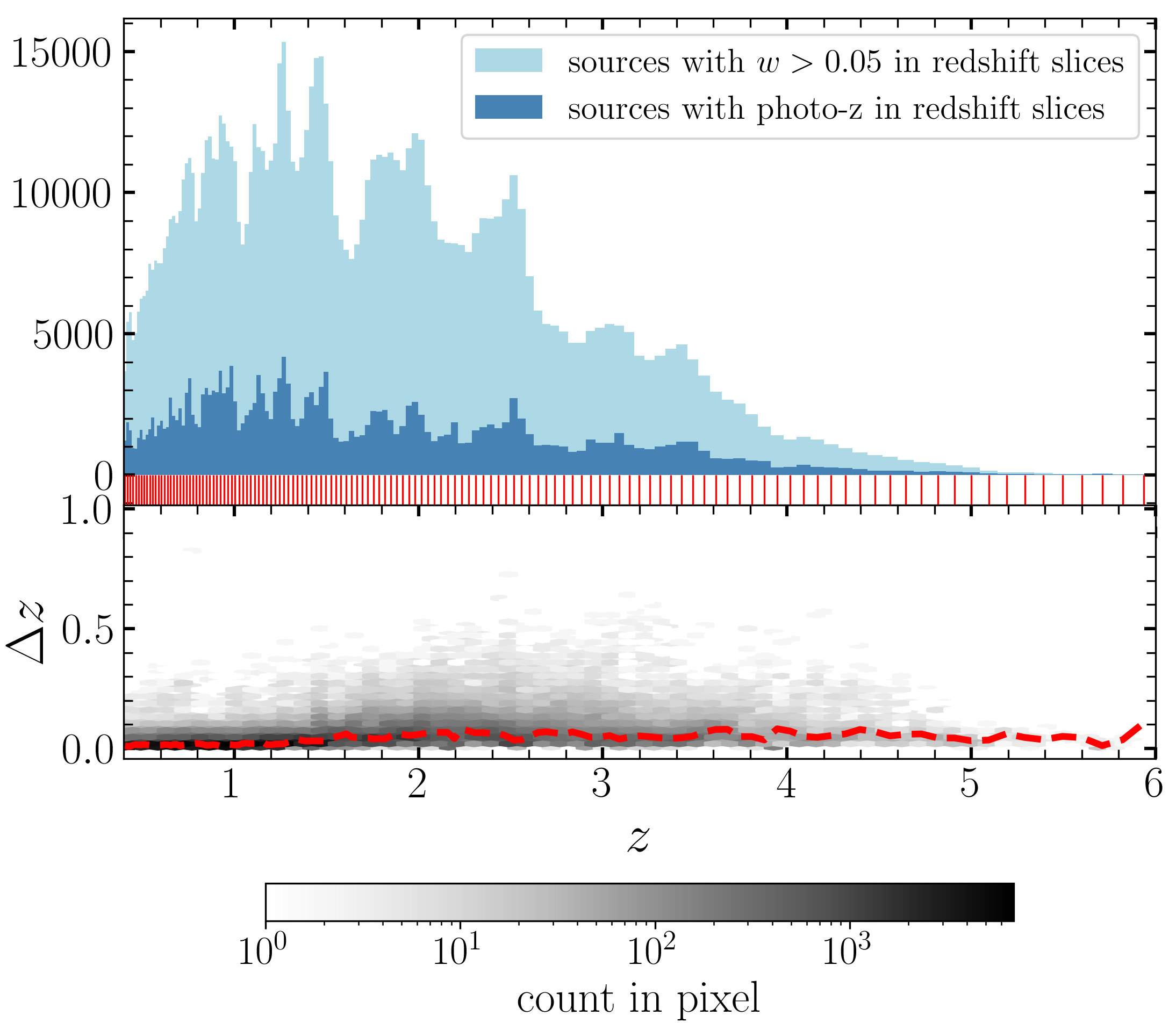}
    \caption{Top: shows the distribution of galaxies across $0.4<z<6$. The dark blue histogram represents the count of galaxies that have photo-$z$ measurement within a redshift slice and the light blue histogram shows the distribution of galaxies that have weights above $w_{th}=0.05$ in each redshift slice. Redshift slice centers are shown as vertical red lines between two panels. Bottom: shows the photo-$z$ uncertainty (half of the $68\%$ confidence interval of photo-$z$) as a function of redshift. The dashed red line depicts the median of the photo-$z$ uncertainties in each redshift slice.}
    \label{histogram}
\end{figure}

\subsection{Redshift Slices}\label{sec: zslice}
To reconstruct the density field at different redshifts we have two options to deal with the uncertainties of photo-$z$s. One is to adopt wide enough, overlapping redshift slices to consider the contribution of galaxies that have large uncertainties on their photo-$z$ and those that are close to the boundaries of each slice. In this approach, the width of slices is chosen based on the redshift uncertainties. For instance, the median of the photo-$z$ uncertainties can be considered as the width of the redshift slices \citep{darvish2015, Scoville2013}. 

\begin{figure*}
    \centering
    \includegraphics[width = 0.8\textwidth]{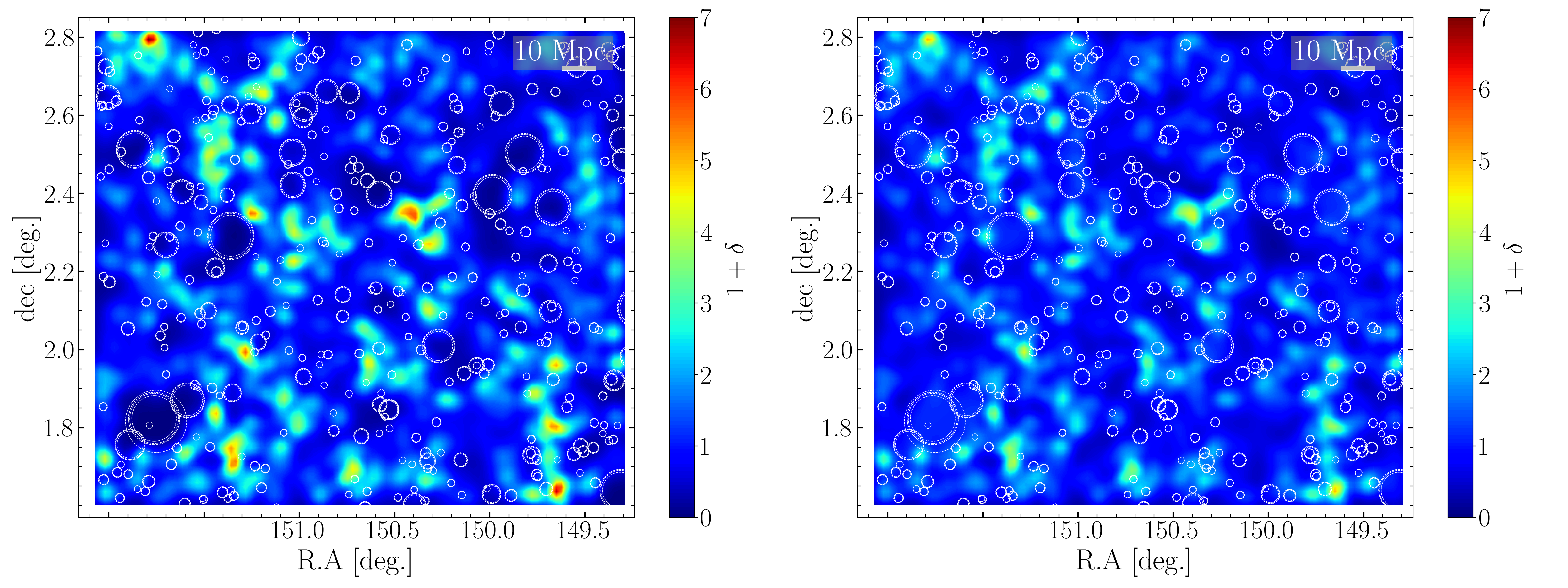}
    \caption{Reconstructed density map at $z = 2.93$. The left panel shows the reconstructed density map without correction for masked regions. The right panel presents the same density map after applying corrections for masked regions by populating them with uniformly distributed artificial sources. This results in a smoother density field, particularly around large masked areas. For easier comparison, some of the largest masks are plotted with white dashed circles.}
    \label{stars}
\end{figure*}

An alternative approach, which is used in this work, is to assign a weight to each galaxy at each redshift slice to incorporate their contribution at all redshifts according to their photo-$z$ PDF \citep{chartab2020}. In this approach, we no longer need to have overlapping slices. We divide our redshift range into slices with a constant comoving width. For that, we need to choose a physically reasonable length scale as the comoving width of redshift slices. This comoving length should be larger than the typical size of structures of our interest (e.g. galaxy clusters), account for the uncertainties in photo-$z$ measurements, and the uncertainty in the redshift direction caused by the peculiar velocity of galaxies in the line of sight, known as redshift space distortion (RSD). Due to RSD, an internal velocity dispersion of $\Delta v$ for a galaxy cluster at redshift z can be translated to a comoving distortion of value $\Delta \chi$ along the line of sight:
\begin{equation}\label{eq: conversion}
    \Delta \chi = \frac{\Delta v}{H_{0}} \frac{1+z}{\sqrt{\Omega_{m0}(1+z)^{3}+\Omega_{\Lambda 0}}}
\end{equation}
This length scale, due to the RSD effect, peaks at $z=(2\Omega_{\Lambda0}/\Omega_{m0})^{1/3}-1 \approx 0.65$: a massive galaxy cluster with internal velocity dispersion $\Delta v \approx 1500 \, km \,s^{-1}$ will be extended $\approx 18 \, h^{-1} \, \text{Mpc}$ in comoving space, at this redshift.

Another constraint in redshift binning is the uncertainty of photo-$z$ measurements. In our sample, a choice of $\Delta \chi = 35 \, h^{-1} \text{Mpc}$ results in relative median uncertainty of value $\Delta z / (1+z) < 0.02 $ in all redshift slices, while it satisfies the minimum required width needed to account for the RSD effect and it is bigger than the typical size of LSS components up to $z_{max} \sim 6$ \citep{Muldrew2015,Chiang2017,ansari2019,zhu2021}.

One can translate this comoving width, $\Delta \chi$, into width of the redshift slices $\Delta z$:
\begin{equation}\label{zbin}
    \Delta z = \frac{\Delta \chi H_{0}}{c} \left[ \Omega_{m} (1+z)^{3} +\Omega_{\Lambda}\right]^{0.5}
\end{equation}
With the choice of $\Delta \chi = 35 \, h^{-1} \text{Mpc}$, we will have 135 redshift slices ranging from $z = 0.4$ to $z = 6$, with slice widths varying from 0.014 (at redshift 0.4) to 0.117 (at redshift 5.936). 

\subsection{Weight Calculation}\label{sec: weightcalculation}
Once we determined the redshift slices we can calculate the weight of a galaxy \textit{g} in the redshift slice \textit{s}, denoted as $w^{g}_{s}$,  which is the probability of galaxy \textit{g} being in the redshift slice \textit{s}. For simplicity, we use a Gaussian probability distribution to calculate these weights for most of the galaxies in the sample that have a single solution for their photo-$z$. We put the center of the Gaussian PDF on the estimated photo-$z$, with a standard deviation calculated using the 68\% confidence interval of photo-$z$. Hence, $w^{g}_{s}$ can be calculated as
\begin{equation}
    w^{g}_{s} = \int_{s} P_{G}(z;\mu = z_{g}, \sigma = \Delta z) \, dz
\end{equation}
where $z_{g}$ is the estimated photo-$z$ and integration domain is over \textit{s}th redshift interval. This is the contribution of galaxy \textit{g} to the density field of \textit{s}th redshift slice (Equation \ref{eq:densitySlice}). For sources that have two peaks in their photo-$z$, a Gaussian PDF is not a good approximation. The fraction of these sources has a monotonically increasing relation with the selected magnitude-cut on the sample. With our choice of magnitude cut ($K_{s}=24.5$), less than 7\% of sources in our sample would have a second solution with $P>5\%$ for their photo-$z$. For these sources, we use their actual photo-$z$ PDF to calculate weights.

Theoretically, all galaxies with a Gaussian probability distribution have non-zero weights in all redshift slices. To reduce the computational time in the density estimation step, we only keep galaxies that have large enough weights, above a weight threshold of value $w_{th} =0.05$, in a redshift slice. Figure \ref{weights} shows the effect of different choices of $w_{th}$ on the sample. The upper panel shows the effective number of galaxies, the summation of all weights in a redshift slice, as a function of redshift. The black curve represents the original sample (without threshold on weights) which is plotted as a reference. A higher threshold decreases the effective number of galaxies in all redshifts. The middle panel shows the effective number of galaxies in each redshift slice divided by the effective number of galaxies in all slices. The lower panel shows the fraction of sources in the sample that enter the next step, or the factor by which the threshold on weights reduces the computation time. The choice of $w_{th}=0.05$ significantly shrinks the sample size in all redshift slices while minimally affecting the distribution of galaxies across the whole redshift range.

The upper panel in Figure \ref{histogram} shows the distribution of galaxies as a function of redshift. The dark blue histogram represents the number of galaxies that have measured photo-$z$ within a redshift slice and the light blue histogram shows the distribution of galaxies that have weights above $w_{th} = 0.05$ in each redshift slice. Vertical red lines between two panels show the centers of 135 redshift slices. The bottom panel in Figure \ref{histogram} shows the uncertainty of photo-$z$s (half of the $68\%$ confidence interval of calculated photo-$z$) as a function of redshift. The red dashed line shows the median of redshift uncertainties in each redshift slice. 

\subsection{Masked Regions} \label{sec:maskedregions}
COSMOS2020 flags objects that are in the regions covered/affected by bright stars in the HSC survey (FLAG\_HSC), and by bright stars in the legacy Supreme-cam data (FLAG\_SUPCAM). These objects are affected by the fluxes of nearby stars or other artifacts. \citep{coupon2018} provide bright star masks from the HSC-SSP PDR2 which is used to flag objects in the vicinity of these sources (red circles in Figure \ref{cosmos}). Moreover, artifacts in the Supreme-Cam images are masked using the same mask as in COSMOS2015 \citep{weaver2022}. Approximately 18\% of sources are located within these masked regions, where measurements (photometry and SED fitting) are not reliable \citep{weaver2022}. Therefore, these sources are not included in our analysis. 

Exclusion of these flagged sources leads the density estimator to underestimate the densities around masked regions. To account for this error, in each redshift slice, we populate the masked regions with a uniform distribution of \enquote{artificial} sources that meet the following two criteria:
\begin{itemize}
    \item The number density of the \enquote{artificial} sources is equal to the average number density of galaxies (actual data) in the field, excluding the masked area:
    \begin{equation}
        \bar{n} = \frac{\sum_{g}w^{g}_{s}}{A_{\text{field}}-A_{\text{masked}}}
    \end{equation}
    Where $A_{\text{field}}$ represents the total area of the field and $A_{\text{masked}}$ refers to the total masked area.
    \item We choose an identical weight for all these artificial sources such that they do not change the average weight of actual galaxies in a redshift slice:
    \begin{equation}
        \bar{w} = \frac{\sum_{g} \, w^{g}_{s}}{N_{g}^{s}}
    \end{equation}
    where $N_{g}^{s}$ is the number of actual galaxies in the $s$th redshift slice. 
\end{itemize}
Figure \ref{stars} presents a comparison between the constructed density field before and after performing corrections for masked regions at $z=2.93$. For easier comparison, some of the largest masks are plotted as white dashed circles. As we can see, the constructed density field around these masked regions will be affected by the lack of sources in these regions if we do not implement corrections.

\begin{figure*}
    \centering
    \includegraphics[width = 1\textwidth]{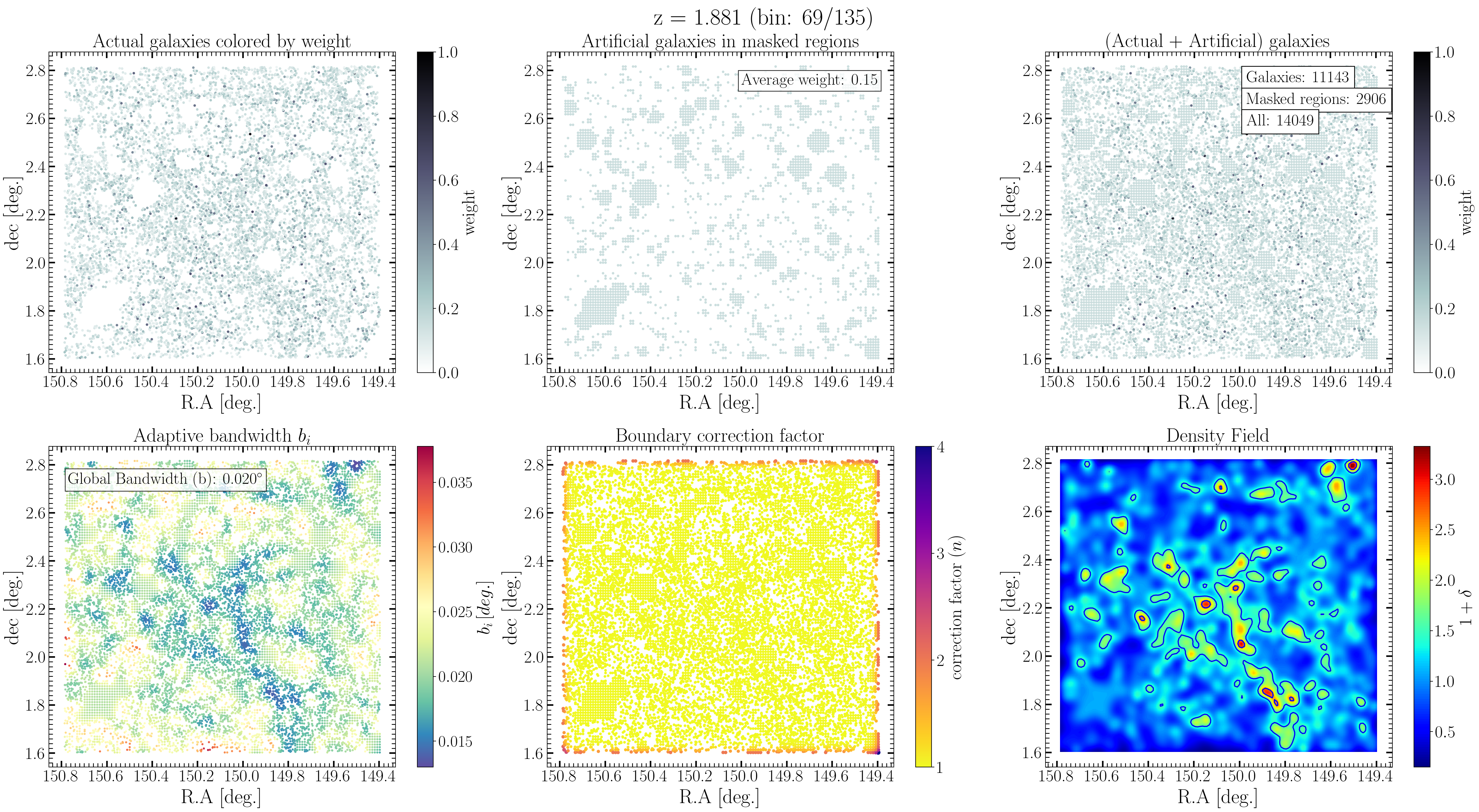}
    \caption{This figure shows the steps taken in the construction of the density map for a redshift slice centered at $z = 1.881$. Upper left: distribution of sources that have weights above $w_{th}$. Upper middle: uniformly distributed artificial sources populated in the masked regions. Upper right: actual galaxies combined with artificial points. Sources in all upper panels are colored by weight. Lower left: actual/artificial sources colored by adaptive bandwidth. Lower middle: the combination of actual/artificial sources colored by the edge correction factor.  Lower right: final density map. Contours are placed at $1+\delta =1.5$ and $2.5$.}
    \label{allsteps}
\end{figure*}

\subsection{Bandwidth Selection}\label{sec:bandwidthselection}
The next step is to choose a bandwidth representing the scale to which the kernel smooths the field, termed \enquote{global bandwidth}. An optimum bandwidth should be chosen based on the number density of sources in each $z$-slice and the extent to which they are clustered: a larger bandwidth results in an over-smoothed field and consequently less information about fine structures while a small bandwidth results in an under-smoothed field with high variance and uncorrelated small scale structures. Selection of the optimum bandwidth is a challenging part of the wKDE. Several methods have been suggested to find the optimum bandwidth. It can be motivated by the physical size of structures of interest in the study. For instance, \citep{darvish2015} adopt a constant global bandwidth of physical length $h = 0.5 \, \text{Mpc}$ for all redshifts, which corresponds to the characteristic size, $R_{200}$, for X-ray clusters and groups in the COSMOS field. \citep{chartab2020} employ the leave-one-out Likelihood Cross-Validation (LCV) method \citep{Hall1982} to find the optimum global bandwidth for each $z$-slice.

As we perform the analysis in a broad redshift range ($0.4<z<6$), adopting a constant physical size for the kernel bandwidth will be an oversimplification and leads us to an unfair estimation of the density field: a fixed bandwidth that performs well at lower redshifts is not our best choice at higher redshifts where the source distribution is more sparse. Therefore, we need to consider the varying number of sources in each redshift slice to set the appropriate bandwidths. We use the LCV method \citep{Hall1982,chartab2020} to find the most likely bandwidth in each $z$-slice, $b_{s}$, given the distribution of sources with specified weights. The method involves a grid search on a range of bandwidths, calculating the likelihood of each candidate bandwidth, and finding the bandwidth that yields the highest likelihood value as the optimal bandwidth. The outcome of this method, $b_{s}$, is data-driven, without any presumption on the bandwidth size, and asymptotically minimizes an integrated squared error in the estimated density:
\begin{equation}\label{equation:LCVformula}
    \text{LCV}(b_{s}) = \frac{1}{N} \sum_{k=1}^{N} \log \sigma_{-k}(\textbf{X}_{k})
\end{equation}
where N is the total number of sources in the field and $\sigma_{-k}(\textbf{X}_{k})$ is the calculated density at position $\textbf{X}_{k}$ leaving the \textit{k}th data point out of the sample (if we do not remove the \textit{k}th data point, the optimal b would be zero).

Next, we determine a local adaptive bandwidth for each point, $b_{i}$, adjusting it according to the clustering level in the surrounding area. In regions with higher clustering, a smaller bandwidth is used for better resolution of smaller structures. Adaptive bandwidth prevents over-smoothing in dense areas and accommodates broader correlations in sparse regions. We calculate $b_{i}$ as \citep{Abramson1982,darvish2015,chartab2020}:
\begin{equation}\label{equation:adaptiveformula}
    b_{i}=b \left( \frac{\sigma(X_{i})}{g} \right)^{-\alpha}
\end{equation}
where $g$ is the geometrical mean of the estimated surface density, $\sigma(X_{i})$, for all sources in the field: 
\begin{equation}
    \log g = \frac{1}{N} \sum_{i}^{N} \log \sigma(X_{i})
\end{equation}
Where $\alpha$ is a constant sensitivity parameter ranging from 0 to 1 and can be determined through simulation. We choose  $\alpha=0.5$, as it has minimal impact on the outcome. 

\subsection{Edge Correction}\label{sec:boundaryeffect}
The wKDE algorithm is effective in areas away from the edges of the field. However, it tends to underestimate density near the edges. This affects only a minor portion of our sources, given the wide area of the COSMOS field. In this section, we implement a correction to address this error. Several methods have been developed to mitigate the edge effect, e.g., the reflection method \citep{schuster1985}, the boundary kernel method \citep{muller1991}, the transformation method \citep{marron1994}, and renormalization method \citep{Jones1993}. In this study, we adopt the re-normalization method: the expectation value of the density field at point $X_{0}$, up to the first order, is 
\begin{equation}
    \mathbb{E} \sim \sigma^{s}_{\text{True}}(X_{0}) \int_{S}K(X_{i};X_{0})
\end{equation}
where $\sigma_{\text{True}}^{s}(X_{0})$ is the true value of the density field at position $X_{0}$ and the integration domain is over the whole field with area S. A reasonable choice for the correction of the edge effect is \citep{chartab2020} 
\begin{equation}
    \sigma^{s}_{\text{corrected}}= \sigma^{j}(X_{0})n(X_{0})
\end{equation}
where $n(X_{0})$ is defined as follows:
\begin{equation}
    n^{-1}(X_{0})=\int_{S}K(X_{i};X_{0})
\end{equation}
The correction factor, $n$, ranges from 1 for a point far from the edges to $\sim 4$ right at one of the corners. 

\begin{table*}[ht]
\caption{Density field measurements for sources in COSMOS2020 catalog.} 
\centering
\begin{tabular}{l c c c c c c c}
\hline\hline 
ID & photo-$z$ & R.A. & Dec. & density contrast  & Comoving density & Physical density & SF/Q\\
   &  & (deg.)  & (deg.) &($\delta$)  & ($\text{Mpc}^{-3}$) & ($\text{Mpc}^{-3}$) &   \\ [0.5ex]
\hline 
964384 & 1.6049 & 149.97682 & 2.4550 & -0.0438 & 0.0011 & 0.020 & 1\\ 
964388 & 1.9782 & 150.44084 & 2.5423 & 0.1876 & 0.0016 & 0.043 & 0\\
964392 & 0.8828 & 149.92698 & 2.3762 & 2.7537 & 0.0317 & 0.213 & 1\\
964393 & 3.4578 & 149.97589 & 2.4575 & 0.6914 & 0.0003 & 0.030 & 1\\
964394 & 0.4715 & 150.22125 & 1.7919 & 0.4254 & 0.0163 & 0.051 & 1\\ [1ex]
\hline 
\end{tabular}
\label{table:catalog}
\vspace{0.15in}
\tablecomments{This table is published in its entirety in the machine-readable format.}
\end{table*}

\subsection{Density Map Construction}
Figure \ref{allsteps} illustrates the complete process of constructing density maps in a redshift slice. Here we summarize the process:
\begin{enumerate}
    \item \textbf{Top-Left.} We calculate the weight for all galaxies in a redshift slice, either assuming a Gaussian zPDF or by using the actual redshift PDF for those that have two solutions for their photo-$z$ with $P>5\%$. 
    \item \textbf{Top-Middle.} We populate the masked regions with artificial sources of uniform distribution and combine them with the actual galaxies (\textbf{Top-Right}). 
    \item \label{bandwidthcalc} \textbf{Bottom-Left.} We find the optimum global bandwidth in each redshift slice, $b(s)$, using the leave-one-out Likelihood cross-validation method (Equation \ref{equation:LCVformula}) and then calculate the adaptive bandwidth (Equation \ref{equation:adaptiveformula}) at the position of each galaxy.
    \item \textbf{Bottom-Middle.} We calculate the correction factor for all sources to compensate for the edge effect using the global bandwidth calculated in step \ref{bandwidthcalc}. 
    \item \textbf{Bottom-Right.} The last step is to construct the over-density maps and calculate the densities for all sources. Over-densities are calculated using the background surface density, $\sigma_{\text{median}}$, defined as the median of the reconstructed surface density field in each redshift slice.  
    \begin{equation}
        1+\delta = \frac{\sigma}{\sigma_{\text{median}}}
    \end{equation}
     With our choice of kernel bandwidth (calculated from LCV), $\sigma_{\text{median}}$, is almost constant at all redshifts. The choice of the median value to define background density minimizes bias resulting from outliers.
\end{enumerate}

\begin{figure*}
    \centering
    \includegraphics[width = 1\textwidth]{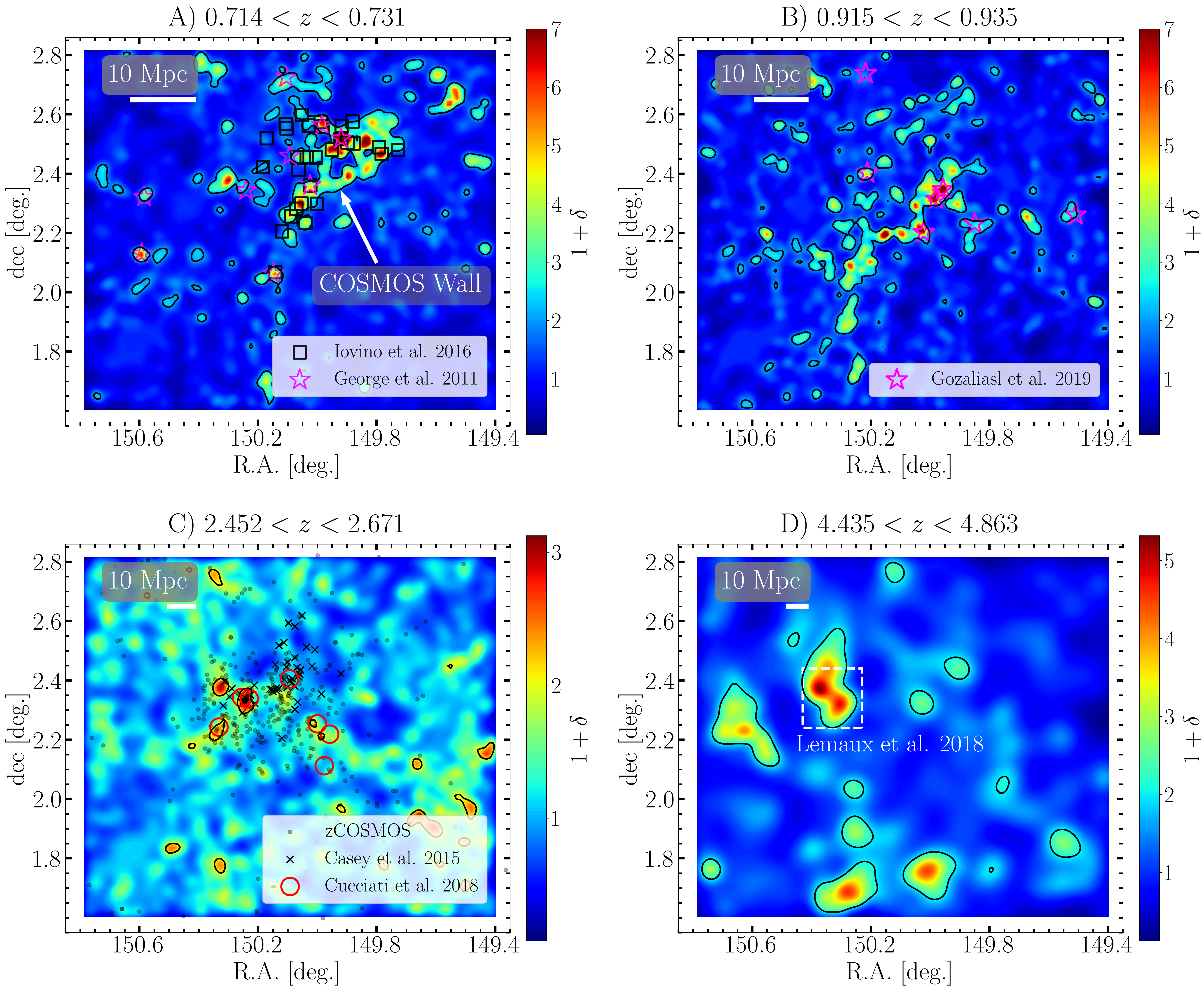}
    \caption{Examples of constructed density maps are shown. A) a filamentary structure, known as COSMOS wall, at $z\sim0.722$ along with galaxy groups reported in \citep{iovino2016} (black boxes) and X-ray groups from \citep{george2011} (magenta stars); B) an elongated structure at $z\sim 0.925$ (can also be found in \citealt{darvish2015}; their Figure 2) along with X-ray galaxy groups within this redshift range from \citep{Gozaliasl2019} shown by magenta stars; C) overdensity map stacked over the range $z\sim 2.452-2.671$ along with confirmed members of PCL1002 \citep{casey_massive_2015} (black crosses), and 7 density peaks in \enquote{Hyperion} proto-supercluster at $z\sim 2.47$ \citep{cucciati_progeny_2018} (red circles). Spectroscopic confirmations from zCOSMOS public catalog (\citealt{lilly2007}, \citealt{lilly2009}, and Khostovan et al. In preparation) are shown with black dots. D) a very distant and massive protocluster, known as \enquote{Taralay} (PCl J1001+0220), at $z\sim 4.57$ studied in \citep{lemaux_vimos_2018,staab2024}. Contours are placed at overdensity $(1+\delta)=2$ in all overdensity maps.}
    \label{fig:densitymaps}
\end{figure*}

Comoving density, $\rho_{\text{comoving}}$, is defined as the number of galaxies in $1 \, \text{Mpc}^{3}$ of comoving space. We calculate the comoving density as $\rho_{\text{comoving}} = (1+ \delta)\bar{\rho}$, where $\bar{\rho}$ is the average number density of galaxies in $s$th redshift slice: 
\begin{equation}
    \bar{\rho} = \frac{\sum_{g}w_{s}^{g}}{\mathbb{V}_{s}}
\end{equation}
Where $\mathbb{V}_{s}$ is the comoving volume associated with the $s$th redshift slice and $\sum_{g}w_{s}^{g}$ is the effective number of galaxies in the selected redshift slice.

Table \ref{table:catalog} presents a portion of the full density catalog, including COSMOS ID, photo-$z$, R.A., and Dec. (from COSMOS catalog); and the measured density contrast, comoving, physical density, and star-forming/quiescent flag (explained in Section \ref{sec:envdependence}). The full electronic density catalog is published in its entirety.

\section{Results \& Discussion} \label{sec:results}
In this section, we present the results of density estimation and utilize them to study the environmental dependence of star formation activity. Our analysis involves two distinct galaxy groups: an overall sample encompassing both star-forming and quiescent galaxies, and a sample of only star-forming galaxies. Additionally, we explore the redshift evolution of this relation by dividing the entire sample into six cosmic epochs (redshift intervals).

\subsection{Large Scale Structures (Density Maps)}\label{sec:LSS}
We release overdensity maps along with the spatial distribution of weighted sources for 135 $z$-slices spanning $0.4<z<6$. The full set of maps is available in animated format. Four examples of overdensity maps along with previously studied/confirmed structures in the same redshift are shown in Figure \ref{fig:densitymaps}. Panel A) shows the filamentary structure at $z\sim0.73$ known as \enquote{COSMOS Wall} along with its galaxy groups studied in \citep{iovino2016} and X-ray groups from \citep{george2011}. Panel B) depicts an elongated structure at $z\sim0.92$ and X-ray galaxy groups at the same redshift from \citep{Gozaliasl2019}. Panel C) shows the stacked overdensity field over $2.452<z<2.671$. This is an interval that includes many massive groups and confirmed sources, some of which are shown in the plot: confirmed members of PCL1002 are shown by black crosses \citep{casey_massive_2015}, 7 massive density peaks in \enquote{Hyperion} massive protocluster by red circles \citep{cucciati_progeny_2018}, and spectroscopy confirmations from zCOSMOS-deep catalog by black dots (\citealt{lilly2007}, \citealt{lilly2009}, and Khostovan et al. In preparation). The very distant and massive protocluster PCL J1001+0220 studied in \citep{lemaux_vimos_2018,staab2024} is shown in panel (D). In all panels, contours are placed at overdensity level $(1+\delta)\sim 2$.

\subsection{Environmental Dependence of SFR and its Redshift Evolution}\label{sec:envdependence}
To study the redshift evolution of SFR/sSFR-density, we divide our sample into mass-complete sub-samples in 6 redshift intervals.
The choice of magnitude cut $K_{s}\sim 24.5$ and other selection criteria introduced in Section \ref{sec:data}, result in a magnitude-limited sample that is distinct from the original COSMOS2020 catalog. In a magnitude-limited sample, the minimum stellar mass we have observations for depends on both redshift and stellar mass-to-light ratio. To obtain mass-complete sub-samples in each redshift interval, we follow the method outlined in \citep{pozeti2010,ilbert2013}. We first re-scale the stellar mass of galaxies to a limiting mass, $M_{\text{lim}}$, which is the mass that a galaxy would have at its redshift if we shift its apparent magnitude to the limiting magnitude of the survey or, in our case, the magnitude-cut $K_{s} \sim 24.5$. All types of galaxies above this mass limit are considered to be brighter than the magnitude cut, and potentially observable. The mass re-scaling relation is $\log (M_{\text{lim}}/M_{\odot}) = \log (M/M_{\odot}) + 0.4 (K_{s} - K_{s, \text{cut}})$, where $M$ is the estimated stellar mass of the galaxies reported by \texttt{The Farmer}-\texttt{LePhare} combination. A constant stellar mass-to-light ratio is presumed in this relation. 

\begin{figure}
    \centering
    \includegraphics[width = 0.95\linewidth]{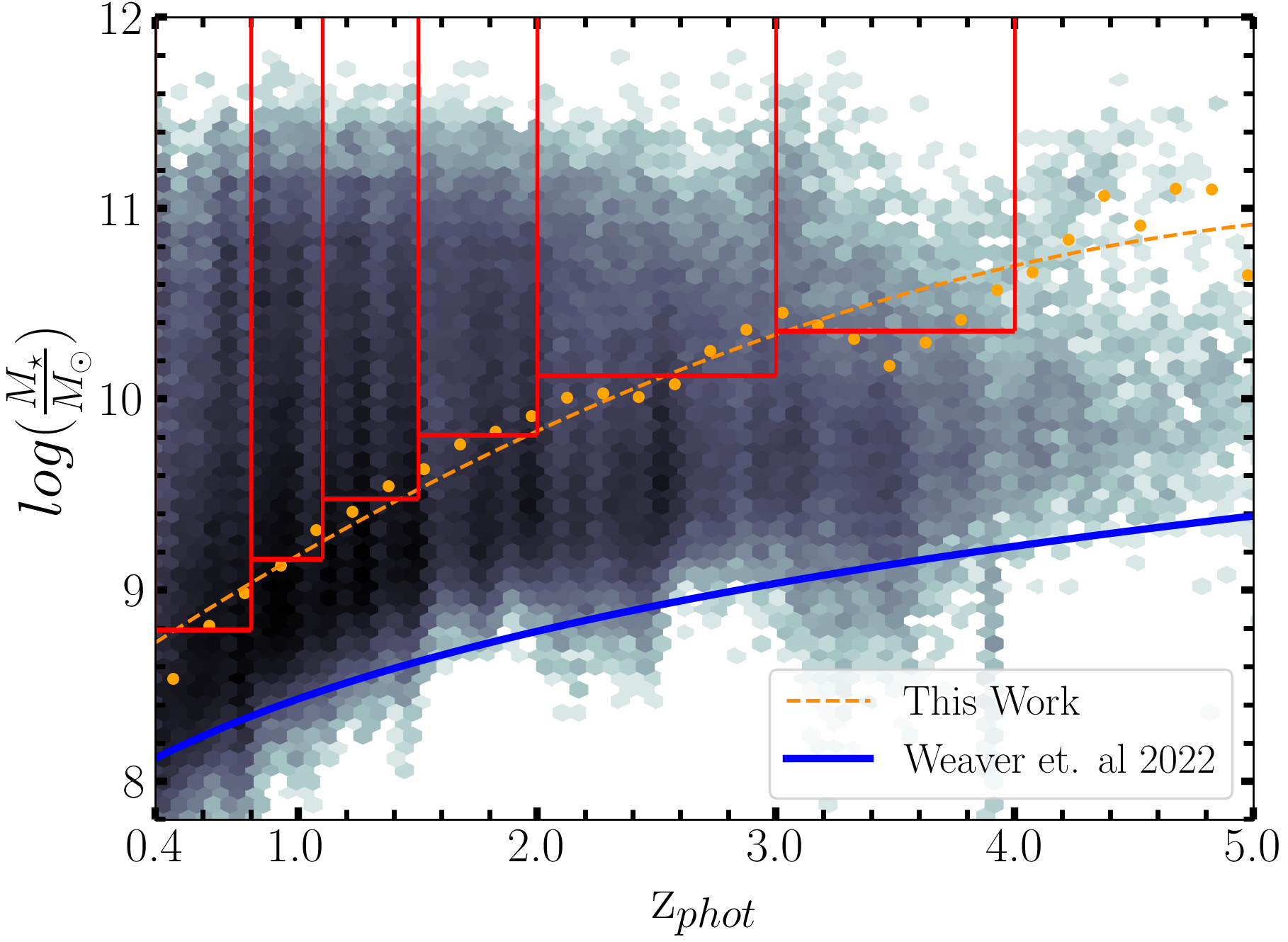}
    \caption{Stellar mass distribution of sources in the sample across the whole redshift range. Red boxes show the boundaries of 6 mass-completeness samples chosen following the method outlined in \citep{pozeti2010}. Properties of these mass-complete samples are summarized in the first three columns of table \ref{table:corrcoeffs}. Orange dots are completeness limits calculated in redshift bins of width $0.1$ and the dashed orange line is the corresponding polynomial fitting function in $(1+z)$. The blue curve is the mass-completeness limit of the original catalog introduced in \citep{weaver2022}, plotted as a reference.}
    \label{masscompleteness}
\end{figure}

At each redshift interval, the final completeness limit $M_{\text{comp}}$, corresponds to the mass below which 95\% of the galaxies' re-scaled masses are populated. This is to ensure that for any subset of galaxies with masses above this limit, not more than 5\% of them could be missed in the lower mass regime. In each interval, less massive (fainter) galaxies that appear at the low redshift end, might be absent at the high redshift end, introducing biases toward more massive galaxies. To minimize this bias, in each redshift interval, we calculate the completeness limit at the high redshift end of the interval. Figure \ref{masscompleteness} shows the distribution of stellar masses versus redshift in our sample. The blue solid line is the mass-completeness fitting function introduced in \citep{weaver2022} for the whole COSMOS2020 catalog plotted here as a reference. Red boxes show the areas that encompass galaxies in the 6 mass-complete bins. Orange dots show the completeness limit we calculated in redshift bins of width 0.1 and the dashed orange line shows the corresponding polynomial fitting function in $(1+z)$.

\begin{figure*}
    \centering
    \includegraphics[width = 1\linewidth]{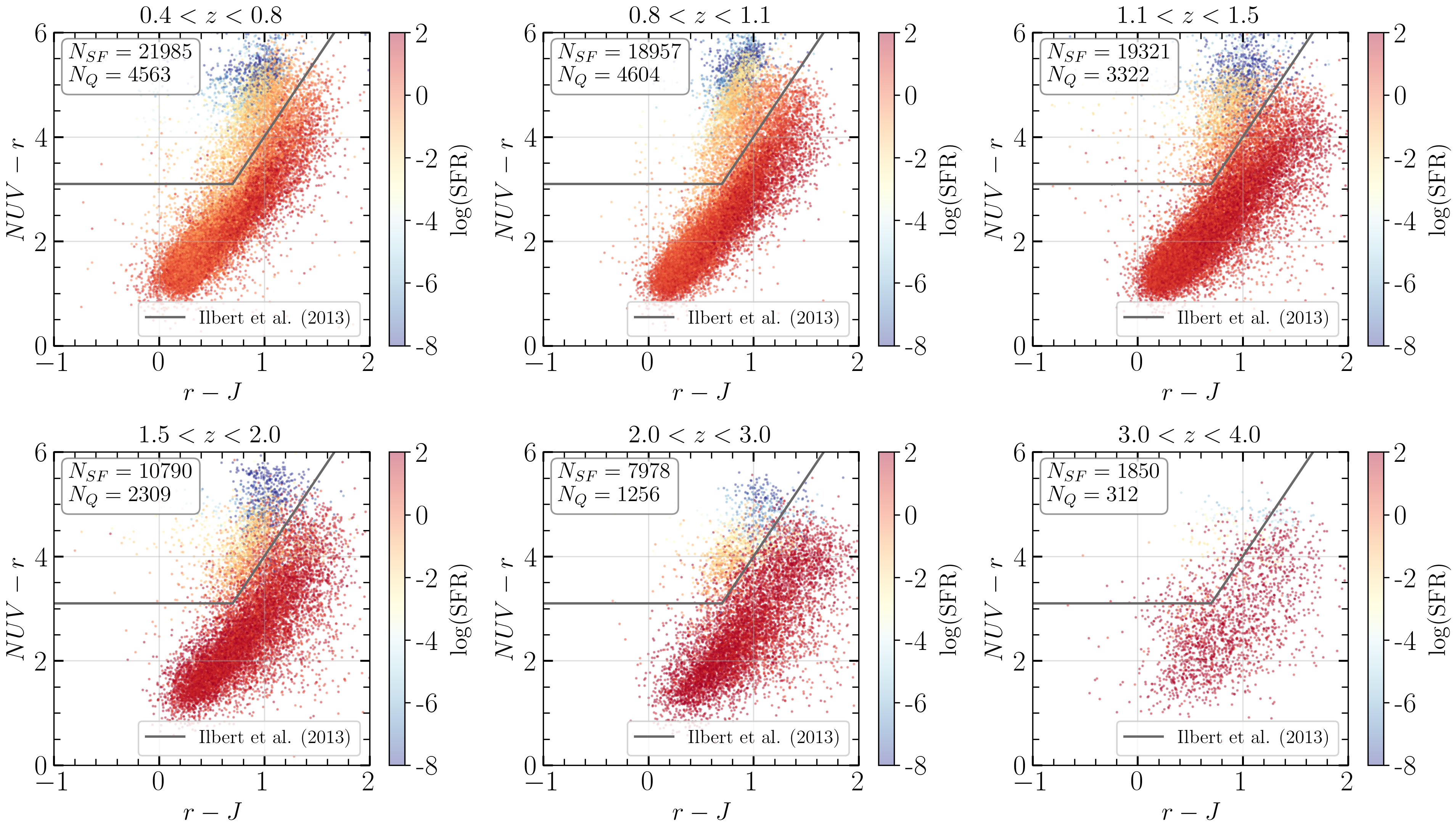}
    \caption{NUV-$r$ vs. $r$-J color-color diagrams and the criteria used to identify quiescent and star-forming galaxies (solid lines) in the 6 mass-complete samples chosen within $0.4<z<4$. Galaxies in the regions defined as NUV-$r>3.1$ and NUV-$r > 3$($r$-J)$+1$ are identified as quiescent \citep{ilbert2013}.}
    \label{figure: NUVrJ}
\end{figure*}

This selection of redshift intervals ensures a substantial number of galaxies in each group. The properties of the resulting sub-samples, including redshift range, mass-completeness limit, and sample sizes are given in the first three columns of Table \ref{table:corrcoeffs}. While we have constructed the density maps for the full redshift range $0.4<z<6$, and we report environmental densities out to $z\sim 5$, we limit the analysis in this part to sources within the range $0.4<z<4$. At $4<z<6$, the number of sources in mass-complete samples becomes sufficiently low ($<300$) to make any statistical conclusion unreliable.

In total, our mass-complete sub-samples include 97,247 galaxies extending up to $z\sim 4$. We then use the rest-frame color-color (NUV-$r$ vs. $r$-J) diagram to identify quiescent galaxies in each bin using the classification criteria introduced by \citep{ilbert2013}: galaxies with a rest-frame color NUV-$r>3.1$ and NUV-$r > 3$($r$-J)$+1$ are flagged as quiescent. Figure \ref{figure: NUVrJ} shows the population of star-forming and quiescent galaxies in each redshift interval all colored by their SFR.

\begin{figure*}
    \centering
    \includegraphics[width = 1\linewidth]{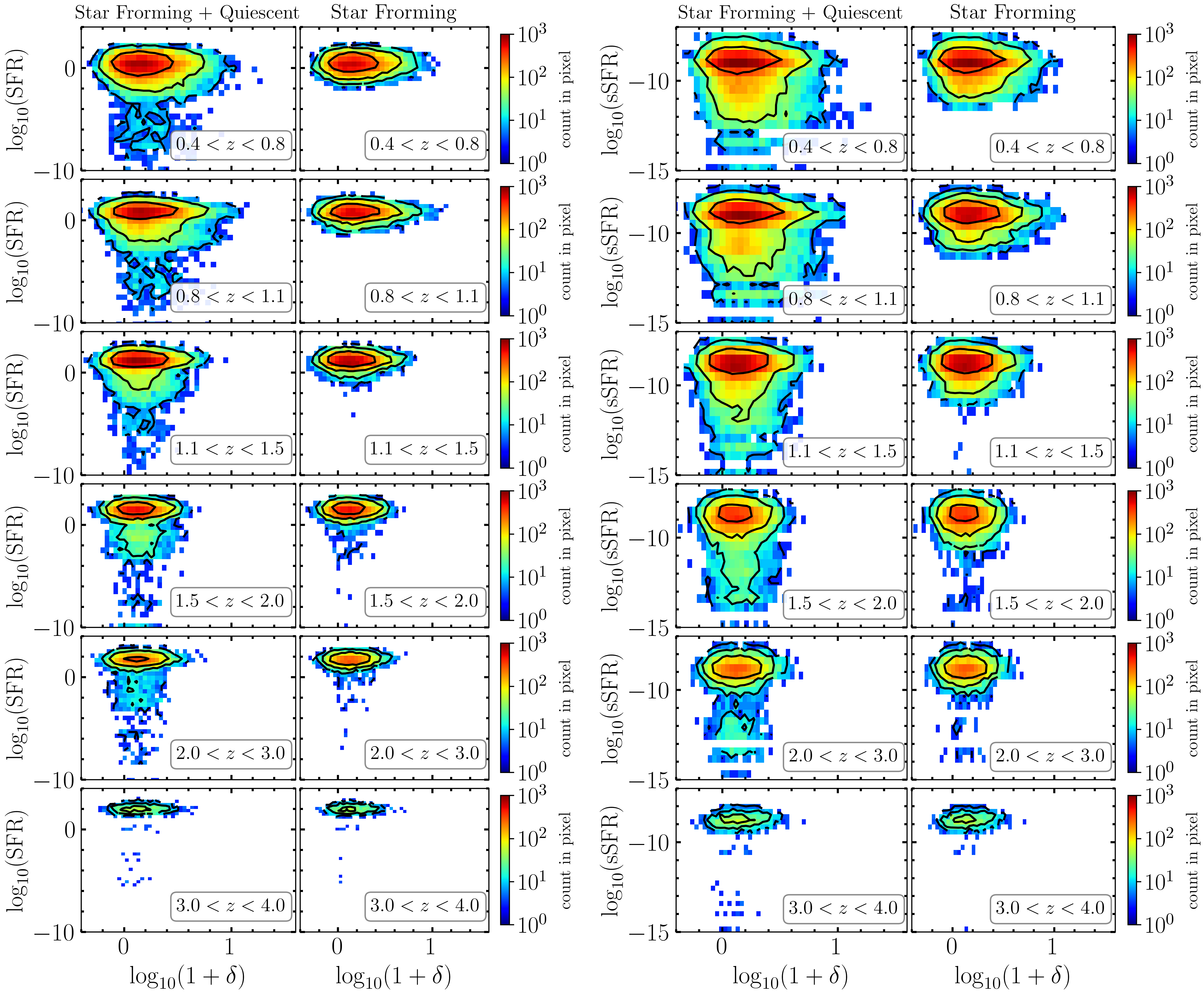}
    \caption{Two columns on the left show SFR vs. environmental density in 6 redshift intervals (column 1: the overall sample; column 2: the star-forming sample). Two columns on the right show sSFR vs. environmental density in 6 redshift intervals (column 3: the overall sample; column 4: the star-forming sample). Colorbars correspond to the population in each pixel and the contours are placed at 3 equally spaced levels between the minimum and maximum population per pixel in each plot and pixels with counts fewer than 5 are set to zero.} 
    \label{fig:SFRsSFR}
\end{figure*}

Figure \ref{fig:SFRsSFR} presents the SFR and sSFR as a function of environmental density for two samples: 1) the overall sample (columns 1 \& 3), and 2) the star-forming sample (columns 2 \& 4). Colorbars correspond to the population in pixels. At lower redshifts ($0.4<z\lesssim1.5$), a significant population of sources is found in high-density environments ($\log (1+\delta) \sim 1$). Conversely, at higher redshifts ($1.5\lesssim z<4$), sources are mostly populated in low/intermediate densities ($\log (1+ \delta) \lesssim 0.6$). In addition, we observed a population of low-SFR/sSFR sources at intermediate densities ($-0.2 \lesssim \log (1+ \delta) \lesssim 0.8$) in columns 1 and 3 (the overall samples). Notably, at $z\lesssim 1.5$, most of these sources vanish in columns 2 and 4, where we exclude quiescent galaxies from our sample. This is attributed to the fact that, among galaxies in low/intermediate density environments, those with low SFR/sSFR are mainly quiescent galaxies. However, this trend lessens at higher redshifts ($1.5<z<4$): there is still a considerable population of \enquote{passive} galaxies at low/intermediate densities ($-0.2 \lesssim \log (1+ \delta) \lesssim 0.8$), even after excluding quiescent galaxies. When interpreting these findings, it is crucial to consider the class imbalance between star-forming and quiescent galaxies at all redshift intervals (Figure \ref{figure: NUVrJ}) and biases toward bright/massive objects at higher redshifts. Furthermore, it should be noted that the COSMOS field, particularly at lower redshifts ($z\lesssim 2$), does not have substantial overdensities compared to other low-redshift surveys or those that are designed to target massive LSS (e.g., SDSS, \citealt{york2000}, EdisCS, \citealt{White2005}; GOGREEN, \citealt{Balogh2017}; ORELSE, \citealt{lubin2009}). As a result, some severe environmental effects, may not be observed within the range of environments we have here ($-0.2\lesssim \log (1+ \delta) \lesssim 1.1$).

To better understand these trends, we present the corresponding binned statistics in Figure \ref{fig: SFRenvdependence}. Average SFR and sSFR are calculated in bins of density, with error bars indicating the standard error of mean values. For both overall and star-forming samples, the SFR-density and sSFR-density trends are almost the same. A considerable difference between SFR-density and sSFR-density dependence is that the average SFR of galaxies increases with increasing redshift, while the average sSFR does not change significantly at different redshifts. The notable decrease in SFR from higher to lower redshifts is related to the decline in the global star formation density of the universe after $z \sim 2-3$ \citep{sobral2013, khostovan2015}. The observed trends for the overall and star-forming samples are discussed in the following sections.

\subsubsection{Overall Sample (Star-Forming and Quiescent)}
The upper panels in Figure \ref{fig: SFRenvdependence} show the average SFR/sSFR as a function of redshift for the \enquote{overall} sample. Since $z\sim4$ to the lowest redshift, the average SFR declines by $1.1$ dex in low-density environments ($\delta \sim -0.5$), while in higher-density environments ($\delta \sim 5$), the SFR decline is approximately $2.8$ dex. The redshift evolution of the sSFR-density relation across the same redshift range is slightly different: at high-density environments ($\delta \sim 3$), there is a significant decline in average sSFR of about $3$ dex, whereas in low-density environments, the sSFR remains almost unchanged. 

Out to $z \sim 1.1$, we see a clear anti-correlation between SFR/sSFR and overdensity. In the lowest redshift bin ($0.4<z<0.8$), SFR decreases by a factor of $\sim 150$ as the density increases from $\delta \sim -0.5$ to $\delta \sim 10$, and in the redshift bin $0.8<z<1.1$, the SFR decreases by a factor of $\sim 10$ across the same range of density. This observed anti-correlation at lower redshifts ($z\lesssim 1$) is well established and in full agreement with many previous studies (e.g., \citealt{patel2009,Scoville2013,darvish2016,Tomczak2019,old2020,chartab2020}). 

At $1.1<z<2$, SFR/sSFR-overdensity correlation weakens such that star formation becomes almost independent of the environment. This observation aligns with findings by \citep{Scoville2013} for $z\gtrsim 1$ (their Figures 15 and 16) and \citep{darvish2016} (Figure 1) for $1.1<z<3.1$, utilizing earlier photometric data from the COSMOS field \citep{ilbert2009, McCracken2012,ilbert2013}. Conversely, some other studies confirm the persistence of the anti-correlation between SFR-density out to $z\sim2$ \citep{Grutzbauch2011,Fossati2017,Ji2018} with \citep{chartab2020} reporting an anti-correlation persisting to even higher redshifts (up to $z\sim 3.5$). We interpret this weakening of correlations as a transitional phase leading to the reversal of trends at higher redshifts.

At $2<z<4$, we observe the reversal of the SFR/sSFR-density relation. For both overall and star-forming samples, at $2<z<3$ and $3<z<4$, SFR increases by a factor of $\sim 10$ as the density increases from $-0.4$ to $5$. The reversal of SFR-density trends at higher redshifts ($z\gtrsim1$) has been reported in several studies \citep{elbaz2007,cooper2008,Santos2015,Welikala2016}. Using a large sample of star-forming H$\alpha$ emitters at $z\sim 0.84$, \citep{sobral2011} reported an increase in SFR of star-forming galaxies at low/intermediate densities, followed by a decline of SFR in the dense environment and clusters. \citep{sobral2011} argued that this might be the reason for inconsistencies at $z \sim 1$, as some studies only reach intermediate/group environments, while others only focus on rich clusters. Using a set of spectroscopic observations \citep{lemaux_vimos_2022} report a positive correlation between the average SFR and galaxy overdensity at ($2<z<5$) across the environmental density range of $-0.2 \lesssim \log (1+ \delta)\lesssim 1$. \citep{lemaux_vimos_2022} showed that the reversal of SFR-overdensity trends persists even when the effect of stellar mass was taken into account.

As we can see, the main controversy centers on trends at higher redshifts ($z\gtrsim 1$). This is partly because, the observed trends are sensitive to larger uncertainties in photo-$z$s at higher redshifts, and redshift binning, which affect the distribution of galaxies in each mass-complete bin. Moreover, some studies attributed the observation of reversal to the effect of cosmic variance in small fields, the small dynamical range of environments (e.g., lack of extremely dense structures in COSMOS field) \citep{sobral2011,Scoville2013,darvish2016} or AGN contamination \citep{Popesso2011}. In addition, different selection functions, especially in studies reliant on spectroscopic samples, introduce further biases (e.g., toward massive clusters or bright sources). Finally, the way the environment is defined/calculated and the statistical interpretation of the results might be another source of inconsistencies between high redshift results. All these factors become more important when interpreting the results in the transition epoch, which occurs between $1<z<2$ according to our findings. A quantitative analysis of these correlations, provided in Section \ref{sec:coefficients}, sheds more light on our interpretation of these trends. 

In addition to the environment, it is shown that stellar mass is another important factor in star formation activity of galaxies \citep{peng2010,Sobral2014,Shivaei2015,Tomczak2019,lemaux2019}. For instance, \citep{darvish2016} find that at a given overdensity, the median SFR for star-forming galaxies is higher for more massive systems out to $z \sim 3$ and \citep{chartab2020} report that the SFR of Massive galaxies ($M\gtrsim10^{11} M_{\odot}$) is inversely correlated at all redshifts and SFR of galaxies with lower stellar mass is almost independent of the environment at $1.2\lesssim z\lesssim 3.5$. Among studies that observe reversal of SFR-density trends, \citep{lemaux_vimos_2022} suggest that high mass galaxies in the denser environments are responsible for this reversal at higher redshifts. We defer a comprehensive analysis of the SFR, stellar Mass, and overdensity relation to our subsequent study (in preparation). However, the correlation between sSFR and overdensity suggests the direct impact of environmental density on the star formation activity of galaxies. 

Despite the ongoing debate around the reversal of the star formation-density trends at higher redshifts, this observation can be attributed to factors such as greater availability of gas supply, tidal interactions, and in general higher star-formation efficiency in high-density environments \citep{wang2018,lemaux_vimos_2022}, potentially leading to enhanced star formation rates at the initial stages of galaxy evolution ($z>2$) for galaxies in rich environments.

\begin{table*}[ht]
\caption{Properties of mass-complete samples and correlation coefficients}
\centering
\begin{tabular}{| l | l l | l l | l @{\extracolsep{35pt}} l |}
\hline\hline
Redshift Range & $\log(\text{M}_{min}/\text{M}_{\odot})$ & Size & \multicolumn{2}{c|}{correlation coeff. (SFR vs. density)} & \multicolumn{2}{c|}{correlation coeff. (sSFR vs. density)} \\
 & & & (SF+Q) & SF & (SF+Q) & SF \\
[0.5ex]
\hline
$0.4<z<0.8$ & 8.791 & 26548 & $-0.060$ & $0.016$ \, \footnotesize(P=0.02) & $-0.097$ & $-0.032$\\ 
$0.8<z<1.1$ & 9.163 & 23561 & $-0.051$ & $0.024$ & $-0.082$ & $-0.023$\\
$1.1<z<1.5$ & 9.479 & 22643 & $-0.007^{\star}$ \footnotesize(P=0.28) & $0.011^{\star}$ \footnotesize(P=0.11) & $-0.021$ & $-0.006^{\star}$ \footnotesize(P=0.37)\\
$1.5<z<2$   & 9.811 & 13099 & $-0.027$ & $0.011^{\star}$ \footnotesize(P=0.24) & $-0.037$ & $-0.001^{\star}$ \footnotesize(P=0.87)\\
$2<z<3$     & 10.122& 9234  &  \ \ $0.062$  & $0.063$ & \ \ $0.062$  & \ \ $0.064$\\
$3<z<4$     & 10.357& 2162  &  \ \ $0.069$  & $0.068$ & \ \ $0.147$  & \ \ $0.151$\\ [1ex]
\hline
\end{tabular}
\label{table:corrcoeffs}
\vspace{0.17in}
\parbox{0.97\linewidth}{\raggedright $^{\star}$ $P>5\%$}
\end{table*}

\subsubsection{Star-Forming Sample}
\begin{figure*}
    \centering
    \includegraphics[width = 1\linewidth]{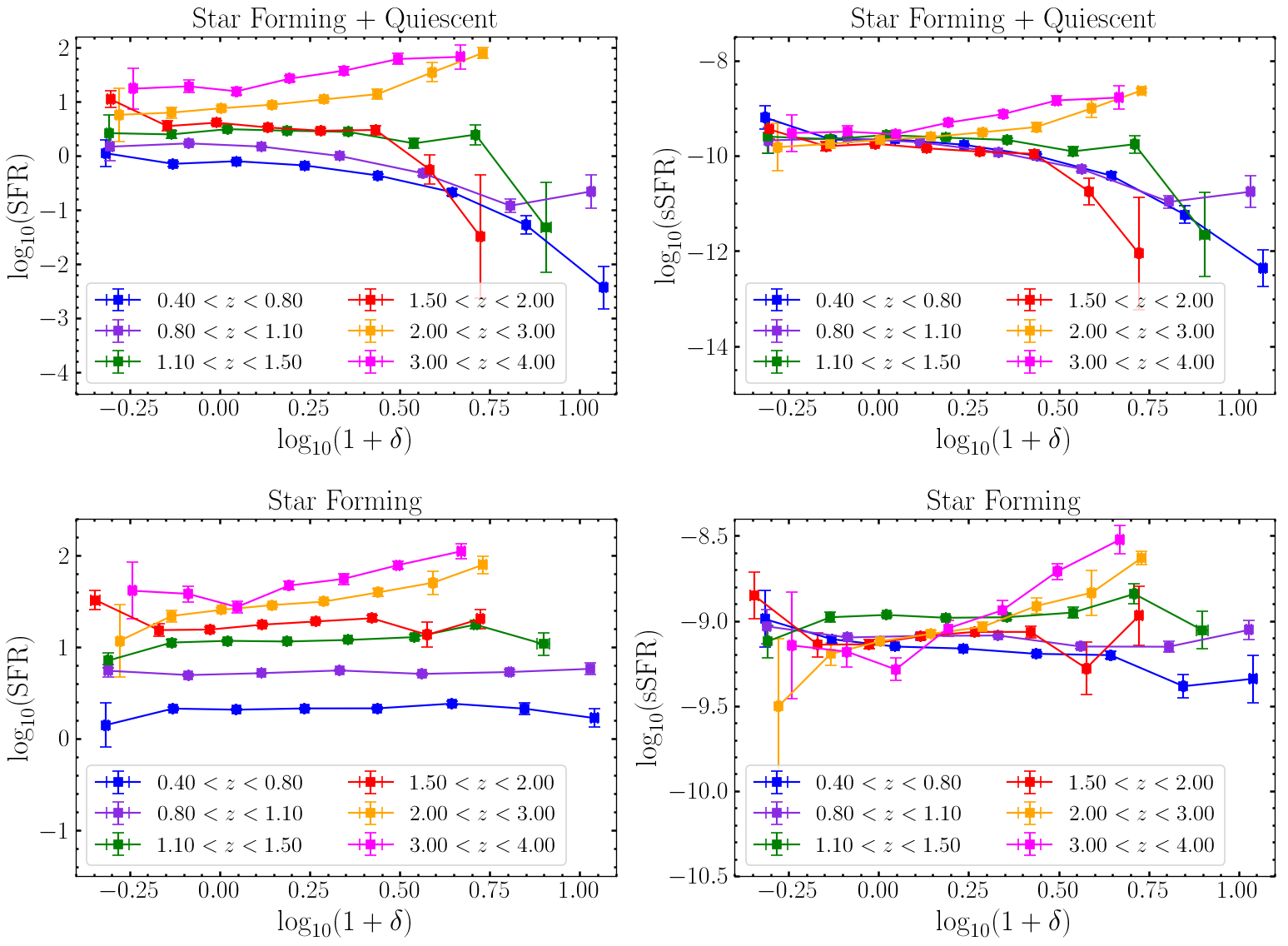}
    \caption{Average SFR (left) and sSFR (right) are plotted versus overdensity for the overall (top) and star-forming samples (bottom) in each redshift interval. Out to $z \sim 1.1$, the median SFR and sSFR decrease with increasing overdensity. At $1.1<z<2$, the SFR and sSFR become almost independent of the environment for both overall and star-forming samples. At higher redshifts ($2<z<4$) there is a positive correlation between star formation activity and the environment for both overall and star-forming samples.}
    \label{fig: SFRenvdependence}
\end{figure*}

The bottom panels in Figure \ref{fig: SFRenvdependence} show the average SFR (bottom left) and average sSFR (bottom right) as a function of overdensity for a sample of star-forming galaxies. At lower redshifts $z\lesssim 1.1$ both SFR and sSFR are almost independent of environment. Therefore, the SFR-overdensity trends observed at $z\sim 1$ in the overall sample are due to the quiescent galaxies that are populated in denser environments. At $1.1<z<2$, both SFR-overdensity and sSFR-overdensity show no significant correlation and trends do not have a monotonic behavior, unlike the lower redshift trends. At our highest redshift bins $2<z<4$, we see a positive correlation in both SFR-overdensity and sSFR-overdensity: SFR increases by a factor of $\sim 8$ as overdensity increases from $-0.5$ to $4.5$. This is a similar behavior to the overall sample, meaning that removing quiescent galaxies does not change the trends at $2<z<4$. This implies a scenario in which a set of processes in high-density environments initially enhance the star formation rate at early epochs ($z\gtrsim 2$). Several processes might contribute to the elevated star formation activity observed in denser environments. These include increased levels of gas accretion, higher merger rates, and the impact of large-scale structures (e.g., tidal effects and galaxy-galaxy interactions). A detailed discussion of these mechanisms can be found in \citep{lemaux_vimos_2022} and references mentioned in their study. In this scenario, the initial enhancement in star formation activity is followed by quenching mechanisms taking effect at $z\lesssim 2$, resulting in anti-correlations between SFR and overdensity observed at lower redshifts. Both quenching mechanisms, environmental quenching (e.g., ram pressure stripping, galaxy-galaxy interactions, galaxy harassment) and stellar mass quenching (e.g., gas outflows and AGN feedback), are shown to be more efficient in denser environments \citep{darvish2016}. Despite these conclusions, we should note that there is an interconnection between massive galaxies and very dense environments causing a degeneracy between stellar mass and environmental density. In other words, the direct role of the environment, regardless of its impact on stellar mass, needs further investigation which is deferred to our future work (Khosravaninezhad et al. in prep). The observed trends between SFR/sSFR and overdensity are discussed quantitatively in the following section.

\subsection{Correlation Coefficients}\label{sec:coefficients}
To quantify our findings in star formation and density relation, we calculated the \enquote{Spearman} correlation coefficient for the reported trends in both overall and star-forming samples. Spearman coefficients determine the degree to which a monotonic function can describe the relationship between two variables \citep{Spearman1904}. Table \ref{table:corrcoeffs} summarizes the properties of mass-complete samples at 6 redshift intervals, along with the Spearman correlation coefficients. Starred coefficients are those that have a $P-$value$>5\%$ and all $P-$values larger than $0.001$ are reported in the table. All other $P$-values associated with the reported coefficients in Table \ref{table:corrcoeffs} are less than $0.001$, suggesting the statistical significance of these trends. For the overall sample, the SFR-density and the sSFR-density relations have generally the same behavior: a negative correlation between SFR/sSFR and density out to $z\sim 2$ (except the $1.1<z<1.5$ bin which has a large $P$-value$=0.28$), which weakens by increasing redshift and is followed by a positive correlation at higher redshifts ($z\gtrsim2$). For the star-forming sample of galaxies, the correlations are generally weaker compared to the overall sample out to $z\sim 2$: all coefficients have an absolute value less than 0.03 with large $P-$values in several redshift bins, suggesting that the reported correlation coefficients are not statistically significant. At the highest redshift bins ($2<z<4$), we see that removing quiescent galaxies does not change the strength and significance of observed trends and we can see a positive relation between SFR/sSFR and density for both overall and star-forming sample.

\section{Summary} \label{sec:Summary}
We use a magnitude-limited ($K_{s}<24.5$) sample of galaxies in the COSMOS2020 catalog to reconstruct density maps across $0.4<z<6$. We choose 135 redshift slices of constant comoving width $35 \, h^{-1} \, \text{Mpc}$ and assign weights to galaxies in all redshift slices using their $z$PDF. We calculate densities adopting the weighted Kernel Density Estimation method, with the choice of von Mises-Fisher kernel, and implement corrections for \enquote{edge effect} and masked regions to improve our estimation of the density field.

We release a publicly available catalog of calculated environmental densities for $\sim200 \, k$ galaxies, along with an animated version of density maps out to $z\sim6$ which can be used to identify LSS. To explore the relation between the star formation activity of galaxies (SFR/sSFR) and environmental density and the evolution of this relation over cosmic time, we provide binned statistics for mass-complete sub-samples within six redshift intervals (details in Table \ref{table:corrcoeffs}). Our findings are summarized as follows:
\begin{enumerate}
    \item In the overall sample, we observe a negative correlation between SFR/sSFR and overdensity at $z \lesssim 1.1$. This correlation diminishes at $1.1<z<2$, and a reversal of trends is noted beyond $z\sim 2$, indicating that galaxies in denser environments exhibit higher star formation activity during the early stages of their evolution.
    \item In the star-forming sample, out to $z\sim 2$, we observe a relatively weak positive (negative) correlation between SFR (sSFR) and overdensity. Beyond $z\sim2$, we observe a positive correlation between SFR/sSFR and overdensity, with the same strength as observed in the overall sample.
    \item At a fixed overdensity, in both overall and star-forming samples, SFR increases as we go to higher redshifts. While we see a similar behavior between sSFR-overdensity at the high-density end of the overall sample, the median sSFR does not change significantly with redshift at lower densities. In the star-forming sample, the average sSFR shows no significant variation with redshift at a given overdensity.
    \item We analyze the strength of these correlations using the Spearman correlation coefficient, the results of which are presented in table \ref{table:corrcoeffs}. The calculated coefficients substantiate the observed trends in Figure \ref{fig: SFRenvdependence}. For the overall sample, we report a negative correlation coefficient between SFR/sSFR and environmental density, diminishing as we go to higher redshifts. This is followed by a shift to a positive correlation at $z\gtrsim 2$, with the transitional phase occurring between $1.1\lesssim z \lesssim 2$.
    \item For the star-forming sample, the coefficients are weaker than those in the overall sample out to $z\sim 2$ and relatively larger $P-$values indicate that the observed trends cannot be confidently explained by monotonic relations, particularly at $1.1<z<2$. At $z>2$ the strength of the observed trends mirrors that of the overall sample, suggesting that excluding the quiescent population does not alter the trends within our highest redshift bins.
\end{enumerate} 

The consistent trends noted in the sSFR-overdensity relation indicate that the mass-normalized star formation activity of galaxies has been affected by their environments since the early phases of galaxy evolution. Increased levels of gas accretion, higher merger rates, and tidal interactions are among the possible environmental processes that enhance star formation activity at early epochs. At lower redshifts, ram pressure stripping, galaxy-galaxy interactions, and suppression of cool gas accretion can be named among the environmental processes that possibly suppress the star formation rate in denser environments. However, it is important to acknowledge that the direct contribution of stellar mass in the evolution of star formation activity needs to be more precisely accounted for. Our ongoing study, building upon this work, is dedicated to a comprehensive analysis of how stellar mass contributes to the trends observed here. This includes assessing the efficiencies of both environmental and mass quenching mechanisms in different environments, and its evolution as a function of redshift.

The accuracy of density estimation methods is heavily dependent on the quality of estimated redshift PDFs. In addition, our statistical conclusions about SFR/sSFR-density relation are limited to the accuracy of SED fitting outputs. Therefore, further improvements in photometry techniques can significantly improve the quality of zPDFs and consequently, SED-derived parameters. Moreover, the accuracy of properties derived from SED-fitting depends on various factors including the dust content of the galaxies. For instance, extremely dusty star-forming galaxies might have their SFR underestimated if their dust obscuration is not adequately accounted for in the SED models. At higher redshifts, particularly around $z\sim2-3$, where the contribution of dusty star-forming galaxies to the overall SFR budget is significant, this bias could potentially impact the observed SFR-density relations.

To have a more accurate understanding of the SFR-environment relation at high redshifts, particularly beyond $z>2$, we still need to acquire deeper observations in larger contiguous fields to improve the completeness of our samples and minimize the effect of cosmic variance. In the near future, the detection of large samples of structures in wide-area surveys such as Euclid, and The Hawaii Two-0 Survey (Zalesky et al. in preparation) is expected to provide further insights into this topic. In a following paper (in preparation), we will study the evolution of star formation activity of galaxies across different components of LSS (field, cluster, filament). We anticipate that with the advent of wide-area surveys in the near future, studies focusing on the influence of large scale structures will become increasingly feasible and relevant.

\section{Acknowledgments}
We are grateful to the anonymous referee for their helpful comments that greatly improved the quality of this work. Some of the data used in this study were obtained at the W.M. Keck Observatory, which is operated as a scientific partnership among the California Institute of Technology, the University of California, and the National Aeronautics and Space Administration. The Observatory was made possible by the generous financial support of the W.M. Keck Foundation. The authors wish to acknowledge the profound cultural role of the Maunakea's summit within the indigenous Hawaiian community. This work is based on data products from observations made with ESO Telescopes at the La Silla Paranal Observatory under ESO program ID 179.A-2005 and on data products produced by CALET and the Cambridge Astronomy Survey Unit on behalf of the UltraVISTA consortium. This work is based in part on observations made with the NASA/ESA Hubble Space Telescope, obtained from the Data Archive at the Space Telescope Science Institute, which is operated by the Association of Universities for Research in Astronomy, Inc., under NASA contract NAS 5-26555. This research has also made use of the zCOSMOS database, operated at CeSAM/LAM, Marseille, France. ST was partially supported by the NSF award 2206813 during this work.

\appendix

\begin{figure}
    \centering
    \includegraphics[width = 0.95\linewidth]{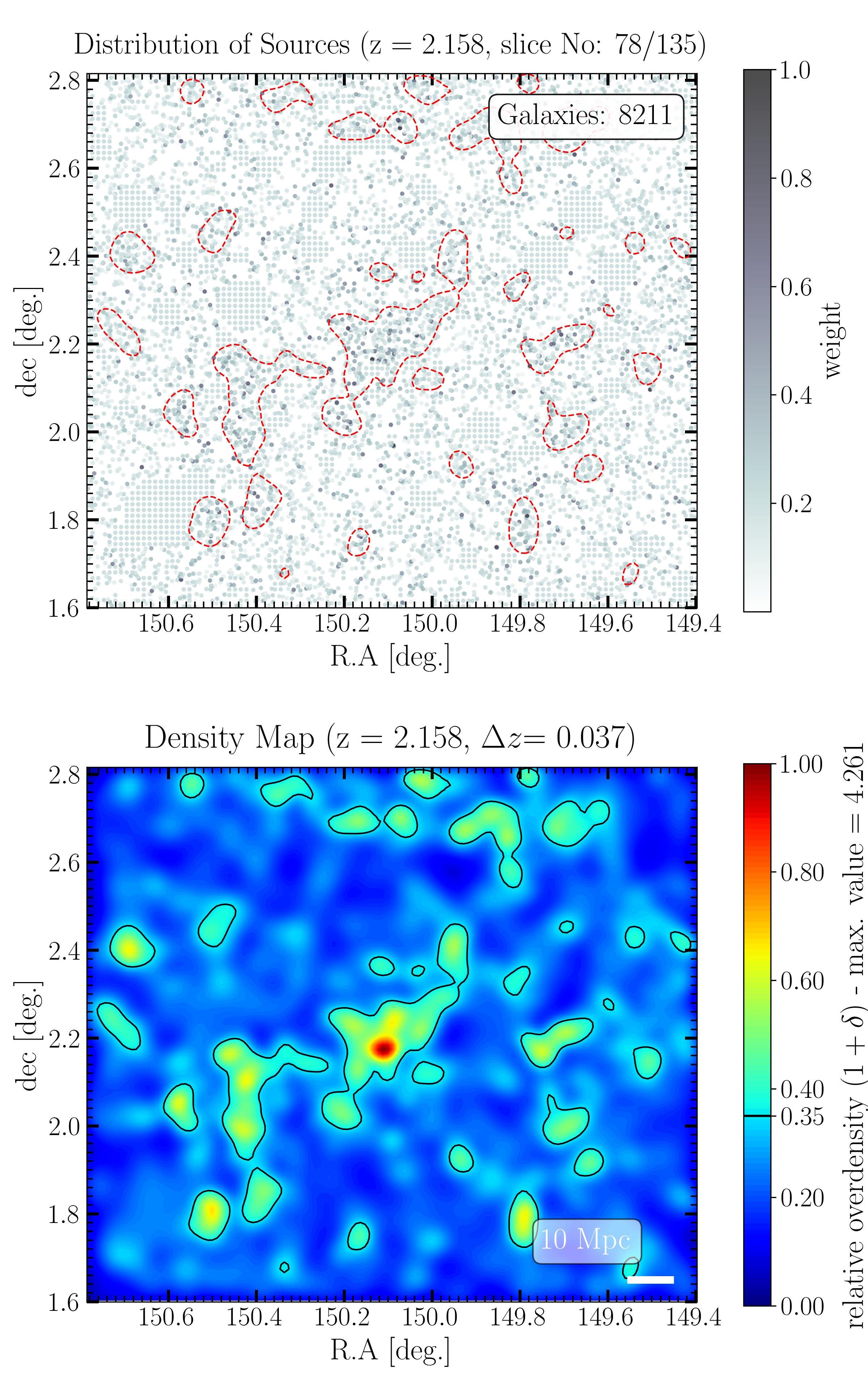}
    \caption{A single frame of the animated density maps which can be found in the HTML version of this article. The top panel shows the spatial distribution of galaxies in the $78$th (out of 135) redshift slice with width $\Delta z \sim 0.037$ and centered at $z\sim 2.158$. Both actual and artificial sources are shown in the top panel and are colored by weight (see Section \ref{sec: weightcalculation}). The bottom panel shows the corresponding density map for the same redshift slice. The density field is clipped and normalized to $1+\delta =8$. When the actual maximum density in a slice exceeds $1+\delta=8$, it is indicated in parentheses in the colorbar label. Contours are placed at the $\sim 0.35$ level of the normalized overdensity in both panels. The animation shows the results for 135 redshift slices, spanning $0.4<z<6$.}
    \label{fig:staticMap}
\end{figure}

\section{Density Maps}\label{sec:appendix}
The full set of density maps for 135 redshift slices spanning $0.4<z<6$ in the COSMOS field can be found in animation format in the HTML version of this paper. As an example, a single frame of this 9-second animation is shown in Figure \ref{fig:staticMap}. Each frame illustrates the distribution of sources, colored by weight, in the top panel. The reconstructed density map for the same redshift slice is shown in the bottom panel. Due to the varying dynamical ranges of environments at each redshift and for better contrast, density maps are clipped and normalized to $1+\delta = 8$. When the actual maximum value exceeds $1+\delta=8$, this value is specifically indicated in parentheses on the colorbar label. Contours are paced at the $0.35$ level of the normalized overdensity.

\bibliography{Ref}{}
\bibliographystyle{aasjournal}
\end{document}